\documentclass[twocolumn, trackchanges]{aastex63}
\usepackage{mathtools}

\newcommand\trg{NGC~4945}
\newcommand\kms{km s$^{-1}$}

\newcommand\mJyb{mJy beam$^{-1}$}
\newcommand\Msun{M$_{\odot}$}
\newcommand\Lsun{L$_{\odot}$}

\defcitealias{Gordon2009}{GS09}

\received{...}
\revised{...}
\accepted{...}
\submitjournal{ApJ}
\shorttitle{Super Star Clusters in \trg}
\shortauthors{Emig et al.}

\turnoffediting

\begin{document}

\title{Super Star Clusters in the Central Starburst of \trg}

%% Use \affiliation for affiliation information. The old \affil is now aliased
%% to \affiliation. AASTeX v6.3 will automatically index these in the header.
%% When a duplicate is found its index will be the same as its previous entry.
%%
%% The new \altaffiliation can be used to indicate some secondary information
%% such as fellowships. This command produces a non-numeric footnote that is
%% set away from the numeric \affiliation footnotes.  NOTE that if an
%% \altaffiliation command is used it must come BEFORE the \affiliation call,
%% right after the \author command, in order to place the footnotes in
%% the proper location.
%%

\correspondingauthor{Kimberly L. Emig}
\email{emig@strw.leidenuniv.nl}

\author[0000-0001-6527-6954]{Kimberly L. Emig}
\affiliation{Leiden Observatory, Leiden University, PO Box 9513, 2300-RA Leiden, the Netherlands}

\author[0000-0002-5480-5686]{Alberto D. Bolatto}
\affiliation{Department of Astronomy and Joint Space-Science Institute, University of Maryland, College Park, MD 20742, USA}

\author[0000-0002-2545-1700]{Adam K. Leroy}
\affiliation{Department of Astronomy, The Ohio State University, 140 West 18th Avenue, Columbus, OH 43210, USA}

\author[0000-0001-8782-1992]{Elisabeth A. C. Mills}
\affiliation{Department of Physics and Astronomy, University of Kansas, 1251 Wescoe Hall Dr., Lawrence, KS 66045, USA}

\author[0000-0002-9165-8080]{Mar\'ia J. Jim\'enez Donaire}
\affiliation{Observatorio Astron\'omico Nacional, Alfonso XII 3, 28014,
Madrid, Spain}

\author[0000-0003-0306-0028]{Alexander G. G. M. Tielens}
\affiliation{Leiden Observatory, Leiden University, PO Box 9513, 2300-RA Leiden, the Netherlands}
\affiliation{Department of Astronomy and Joint Space-Science Institute, University of Maryland, College Park, MD 20742, USA}

\author[0000-0001-6431-9633]{Adam Ginsburg}
\affiliation{Department of Astronomy, University of Florida, PO Box 112055, USA}

\author[0000-0001-9300-354X]{Mark Gorski}
\affiliation{Chalmers University of Technology, Gothenburg, Sweden}

\author[0000-0003-1104-2014]{Nico Krieger}
\affiliation{Max-Planck-Institut f\"ur Astronomie, K\"onigstuhl 17, 69120 Heidelberg, Germany}

\author[0000-0003-2508-2586]{Rebecca C. Levy}
\affiliation{Department of Astronomy and Joint Space-Science Institute, University of Maryland, College Park, MD 20742, USA}

\author[0000-0001-9436-9471]{David S. Meier}
\affiliation{Department of Physics, New Mexico Institute of Mining and Technology, 801 Leroy Pl., Socorro, NM, 87801, USA}
\affiliation{National Radio Astronomy Observatory, P. O. Box O, 1003 Lopezville Rd., Socorro, NM, 87801, USA}

\author{J\"urgen Ott}
\affiliation{National Radio Astronomy Observatory, P. O. Box O, 1003 Lopezville Rd., Socorro, NM, 87801, USA}

\author[0000-0002-5204-2259]{Erik Rosolowsky}
\affiliation{Department of Physics, 4-183 CCIS, University of Alberta, Edmonton, Alberta T6G 2E1, Canada}

\author[0000-0003-2377-9574]{Todd A.\ Thompson}
\affiliation{Department of Astronomy, The Ohio State University, 140 West 18th Avenue, Columbus, OH 43210, USA}

\author[0000-0002-3158-6820]{Sylvain Veilleux}
\affiliation{Department of Astronomy and Joint Space-Science Institute, University of Maryland, College Park, MD 20742, USA}

%%%%%%%%%%%%%%%%%%%%%%%%%%%%%%%%%%%%%%%%

\begin{abstract}

\trg\ is a nearby (3.8~Mpc) galaxy hosting a nuclear starburst and Seyfert Type 2 AGN. We use the Atacama Large Millimeter/submillimeter Array (ALMA) to image the 93~GHz (3.2~mm)  free-free continuum and hydrogen recombination line emission (H40$\alpha$ and H42$\alpha$) at 2.2~pc (0.12\arcsec) resolution. Our observations reveal 27 bright, compact sources with FWHM sizes of 1.4--4.0~pc, which we identify as candidate super star clusters. Recombination line emission, tracing the ionizing photon rate of the candidate clusters, is detected in 15 sources, 6 of which have a significant synchrotron component to the 93~GHz continuum. Adopting an age of $\sim$5~Myr, the stellar masses implied by the ionizing photon luminosities are  $\log_{10}$($M_{\star}$/\Msun) $\approx$ 4.7--6.1. We fit  a slope to the cluster mass distribution and find $\beta = -1.8 \pm 0.4$. The gas masses associated with these clusters, derived from the dust continuum at 350~GHz, are typically an order of magnitude lower than the stellar mass. These candidate clusters appear to have already converted a large fraction of their dense natal material into stars and, given their small free-fall times of $\sim$0.05~Myr, are surviving an early volatile phase. We identify a point-like source in 93~GHz continuum emission which is presumed to be the AGN. We do not detect recombination line emission from the AGN and place an upper limit on the ionizing photons which leak into the starburst region of $Q_0 < 10^{52}$~s$^{-1}$.

\end{abstract}

\keywords{galaxies: individual (\trg) -- galaxies: ISM -- galaxies: starburst -- galaxies: star clusters: general -- galaxies: star formation}

%%%%%%%%%%%%%%%%%%%%%%%%%%%%%%%%%%%%%%%%

\section{Introduction} \label{sec:intro}

Many stars form in clustered environments \citep{Lada2003,Kruijssen2012}. Bursts of star formation with high gas surface density produce massive ($>10^5$~\Msun), compact \citep[FWHM size of 2-3~pc;][]{Ryon2017} clusters, referred to as super star clusters. Super star clusters likely have high star-formation efficiencies \edit1{\citep{Goddard2010, Ryon2014, Adamo2011,Adamo2015, Chandar2017, Johnson2016, Ginsburg2018b}}. They may represent a dominant output of star-formation during the peak epoch of star-formation \citep[$z \sim 1-3$;][]{Madau2014}. The process by which these massive clusters form now may also relate to the origin of globular clusters.

The earliest stages of cluster formation are the most volatile and currently, unconstrained \edit1{\citep{Dale2015, Ginsburg2016, Krause2016, Li2019, Krause2020}}. Characterizing properties of young ($<$10~Myr) clusters is a key step towards understanding their formation, identifying the dominant feedback processes at each stage of cluster evolution, determining which clusters survive as gravitationally bound objects, and linking all of these processes to the galactic environment.

While young clusters of mass $\sim$10$^4$~\Msun\ are found within our Galaxy \citep{Bressert2012,Longmore2014,Ginsburg2018a}, the most massive, young clusters in the local universe are often found in starbursting regions and merging galaxies \edit1{\citep[e.g.,][]{Zhang1999, Whitmore2010, Linden2017}}. Direct optical and even near-infrared observations of forming clusters are complicated by large amounts of extinction. Analyses of optically thin free-free emission and long wavelength hydrogen recombination lines of star clusters offer an alternative, extinction-free probe of the ionizing gas surrounding young star clusters \edit1{\citep{Condon1992a, Roelfsema1992, Murphy2018a}}. However, achieving a spatial resolution matched to the size of young clusters $\mathcal{O}$(1~pc) \citep{Ryon2017} in galaxies at the necessary frequencies and sensitivities has only recently become possible thanks to the Atacama Large Millimeter/submillimeter Array (ALMA).

We have recently analyzed forming super star clusters in the central starburst of the nearby (3.5~Mpc) galaxy NGC~253 at $\sim$2~pc resolution \citep{Leroy2018,Mills2020}. \trg\ is the second object we target in a campaign to characterize massive star clusters in local starbursts with ALMA. 

\trg\ is unique in that it is one of the closest galaxies \citep[$3.8 \pm 0.3$~Mpc;][]{Karachentsev2007} where a detected AGN and central starburst coexist. In the central $\sim$200 pc, the starburst dominates the infrared luminosity and ionizing radiation \citep{Spoon2000, Marconi2000}, and an outflow of warm ionized gas has been observed \citep{Heckman1990, Moorwood1996, Mingozzi2019}. Individual star clusters have not previously been observed in \trg, due to the high extinction at visible and short IR wavelengths \citep[e.g., $A_{\mathrm{V}} \gtrsim 36$~mag; ][]{Spoon2000}. Evidence for a Seyfert AGN comes from strong, variable X-ray emission, as \trg\ is one of the brightest sources in the X-ray sky and has a Compton thick column density of $3.8 \times 10^{24}$~cm$^{-2}$ \citep{Marchesi2018}. A kinematic analysis of H$_2$O maser emission yields a black hole mass of $1.4 \times 10^6$~\Msun\ \citep{Greenhill1997}. 

In this article, we use ALMA to image the 93~GHz free-free continuum and hydrogen recombination line emission (H40$\alpha$ and H42$\alpha$) at 2.2~pc (0.12\arcsec) resolution. This emission allows us to probe photo-ionized gas on star cluster scales and thereby trace ionizing photon luminosities. We identify candidate star clusters and estimate properties relating to their size, ionizing photon luminosity, stellar mass, and gas mass.

Throughout this article, we plot spectra in velocity units with respect to a systemic velocity of $\mathrm{v_{sys}} = 580$~\kms\ in the local standard of rest frame; estimates of the systemic velocity vary by $\pm 25$~\kms\ \citep[e.g.,][]{Henkel2018, Chou2007, Roy2010}. At the distance of 3.8~Mpc, 0.1\arcsec\ corresponds to 1.84~pc.

%%%%%%%%%%%%%%%%%%%%%%%%%%%%%%%%%%%%%%%%

\section{Observations}    \label{sec:obs}

We used the ALMA Band 3 receivers to observe \trg\ as part of the project 2018.1.01236.S (PI: A. Leroy). We observed \trg\ with the main 12~m array telescopes in intermediate and extended configurations. 
Four spectral windows in Band 3 -- centered at 86.2, 88.4, 98.4, and 100.1 GHz -- capture the millimeter continuum primarily from free-free emission and cover the hydrogen recombination lines of principal quantum number (to the lower state) $\mathsf{n} = 40$ and $\mathsf{n} = 42$ from the $\alpha$ ($\Delta \mathsf{n} = 1$) transitions. The rest frequency of H$40\alpha$ is 99.0230~GHz and of H$42\alpha$ is 85.6884~GHz.
In this article, we focus on the 93~GHz ($\lambda~\sim~3.2$~mm) continuum emission and the recombination line emission arising from compact sources in the starbursting region. We image the data from an 8~km extended configuration, which are sensitive to spatial scales of 0.07\arcsec--6\arcsec\ (2--100~pc), in order to focus on the compact structures associated with candidate clusters. We analyze the observatory-provided calibrated visibilities using version 5.4.0 of the Common Astronomy Software Application \citep[CASA;][]{McMullin2007}.

When imaging the continuum, we flag channels with strong spectral lines. Then we create a continuum image using the full bandwidth of the line-free channels. We also make continuum images for each spectral window. For all images, we use Briggs weighting with a robust parameter of $r=0.5$. 

When imaging the two spectral lines of interest, we first subtract the continuum in $uv$ space through a first order polynomial fit. Then, we image by applying a CLEAN mask (to all channels) derived from the full-bandwidth continuum image. Again, we use Briggs weighting with a robust parameter of $r=0.5$, which represents a good compromise between resolution and surface brightness sensitivity. 

After imaging, we convolved the continuum and line images to convert from an elliptical to a round beam shape.  For the full-bandwidth continuum image presented in this article, the fiducial frequency is $\nu = 93.2$ GHz and the final full-width half maximum (FWHM) beam size is $\theta = 0.12$\arcsec. The rms noise away from the source is $\approx$0.017~\mJyb, equivalent to 0.2~K in Rayleigh-Jeans brightness temperature units. Before convolution to a round beam, the beam had a major and minor FWHM of 0.097\arcsec~$\times$~0.071\arcsec. 

For the H40$\alpha$ and H42$\alpha$ spectral cubes, the final FWHM beam size is 0.20\arcsec, convolved from 0.097\arcsec~$\times$~0.072\arcsec\ and from 0.11\arcsec~$\times$~0.083\arcsec, respectively.  The slightly lower resolution resulted in more sources with significantly detected line emission. We boxcar smoothed the spectral cubes from the native 0.488 MHz channel width to 2.93 MHz. The typical rms in the H40$\alpha$ cube is 0.50~\mJyb\ per 8.9~\kms~channel. The typical rms in the H42$\alpha$ cube is 0.48~\mJyb\ per 10.3~\kms~channel.

As part of the analysis, we compare the high resolution data with observations taken in a 1~km intermediate configuration as part of the same observing project. \edit1{We use the intermediate configuration data to trace the total recombination line emission of the starburst.} We use the continuum image provided by the observatory pipeline, which we convolve to have a circular beam FWHM of 0.7\arcsec; the rms noise in the full-bandwidth image is 0.15~\mJyb. The spectral cubes have typical rms per channel of 0.24~\mJyb\ with the same channel widths as the extended configuration cubes. We do not jointly image the configurations because our main science goals are focused on compact, point-like objects. The extended configuration data on their own are well suited to study these objects and any spatial filtering of extended emission will not affect the analysis.

% 350 GHz data from archive
We compare the continuum emission at 3~mm with archival ALMA imaging of the $\nu = 350$~GHz ($\lambda \sim 850~\mu$m) continuum (project 2016.1.01135.S, PI: N. Nagar). At this frequency, dust emission dominates the continuum. We imaged the calibrated visibilities with a Briggs robust parameter of $r=-2$ (towards uniform weighting), up-weighting the extended baselines to produce a higher resolution image, suitable for comparison to our new Band 3 data. We then convolve the images to produce a circularized beam, resulting in a FWHM resolution of 0.12\arcsec\ (from an initial beam size of 0.10\arcsec~$\times$~0.064\arcsec), exactly matched to our 93~GHz continuum image. These data have an rms noise of 0.7 \mJyb\ (0.2~K).

% 2.3 GHz data 
We compare the ALMA data with Australian Long Baseline Array (LBA) imaging of $\nu = 2.3$~GHz continuum emission \citep{Lenc2009}. At this frequency and resolution, the radio continuum is predominantly synchrotron emission. We use the Epoch 2 images (courtesy of E. Lenc) that have a native angular resolution slightly higher than the 3~mm ALMA data, with a beam FWHM of 0.080\arcsec~$\times$~0.032\arcsec\ and an rms noise of 0.082~\mJyb.

%%%%%%%%%%%%%%%%%%%%%%%%%%%%%%%%%%%%%%%%

\section{Continuum Emission}   \label{sec:cont}

\begin{figure}
    \centering
    \includegraphics[width=0.47\textwidth]{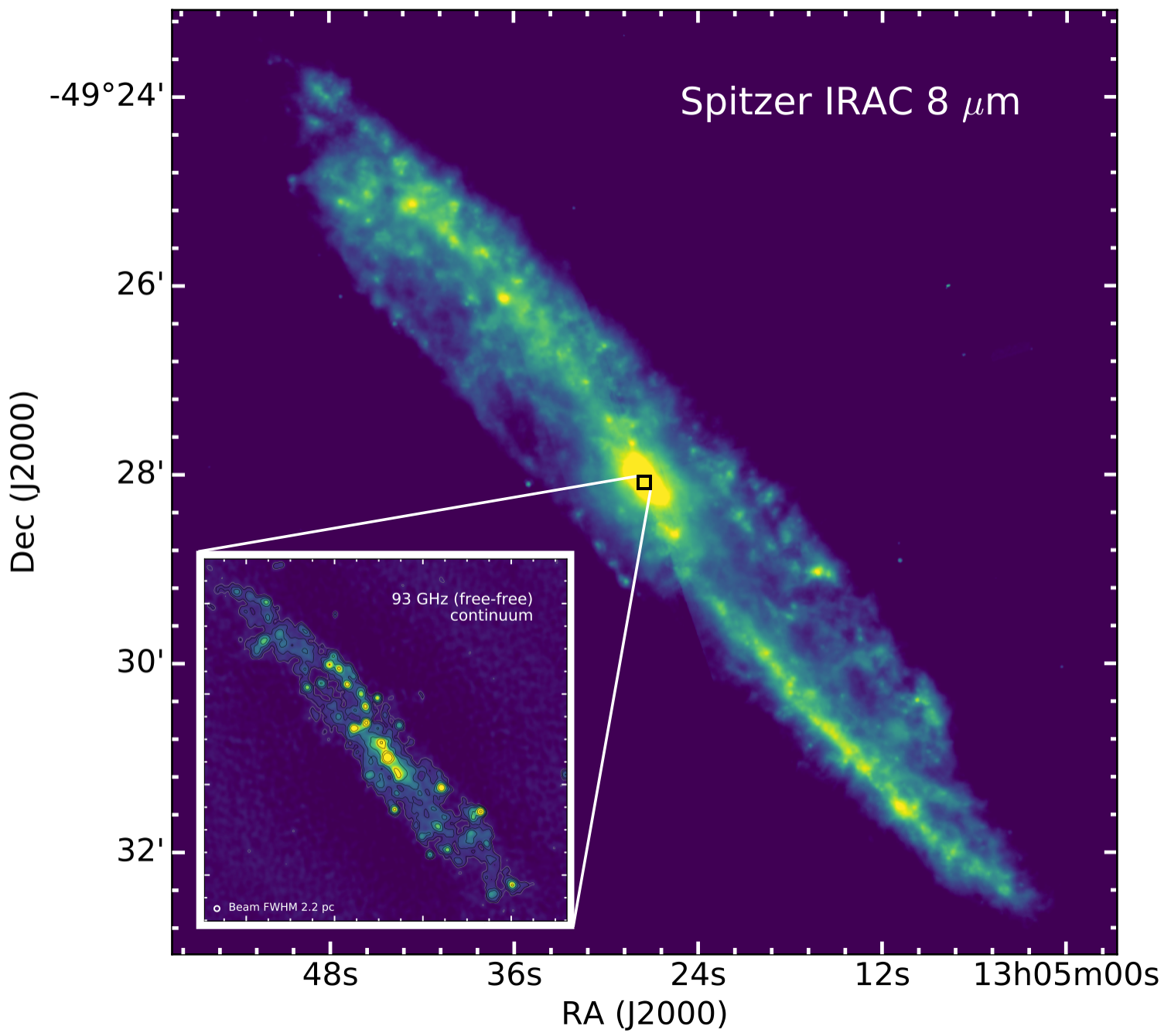}
    \caption{\textit{Spitzer} IRAC 8~$\mu$m emission from UV-heated PAHs over the full galactic disk of \trg. The black square box indicates the $8'' \times 8''$ (150~pc~$\times$~150~pc) central starburst region of interest in this article; the inset shows the ALMA 93~GHz continuum emission.}
    \label{fig:irac}
\end{figure}

The whole disk of \trg, as traced by \textit{Spitzer} IRAC 8~$\mu$m emission (Program 40410, PI: G. Rieke), is shown in Figure~\ref{fig:irac}. The 8~$\mu$m emission predominantly arises from UV-heated polycyclic aromatic hydrocarbons (PAHs), thus tracing the interstellar medium and areas of active star formation. The square box indicates the 8\arcsec~$\times$~8\arcsec\ (150~pc~$\times$~150~pc) starburst region that is of interest in this article.

\begin{figure*}
    %\centering
    \includegraphics[width=0.9\textwidth]{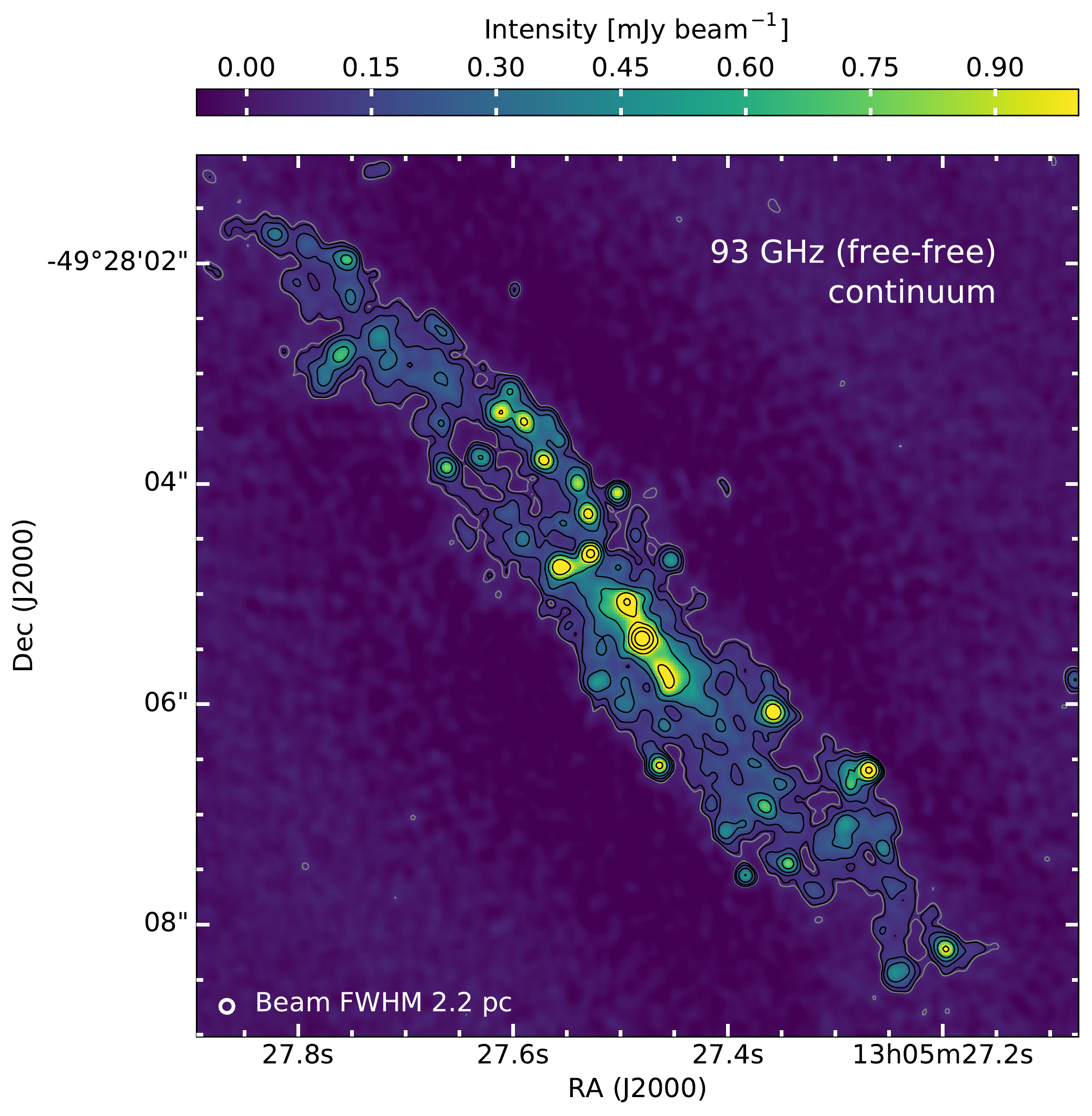}
    \caption{ALMA 93~GHz ($\lambda \sim 3.2$~mm) continuum emission in the central starburst of \trg. The continuum at this frequency is dominated by ionized, free-free emitting plasma. In this paper, we show that the point-like sources are primarily candidate, massive star clusters. The brightest point source of emission at the center is presumably the Seyfert AGN. The rms noise away from the source is $\sigma \approx 0.017$~\mJyb\ and the circularized beam FWHM is 0.12\arcsec\ (or 2.2~pc at the distance of \trg). Contours of the continuum image show $3\sigma$ emission (gray) and $[4,8,16,...256]\sigma$ emission (black).}
    \label{fig:cont93}
\end{figure*}

% 93 GHz emission continuum, Fig 1
Figure~\ref{fig:cont93} shows the 93~GHz ($\lambda \sim 3$~mm) continuum emission in the central starburst of \trg.  Our image reveals $\sim$30 peaks of compact, localized emission with peak flux densities 0.6--8~mJy (see Section~\ref{sec:ps_id}). On average, the continuum emission from \trg\ at this frequency is dominated by thermal, free-free (bremsstrahlung) radiation \citep{Bendo2016}. Free-free emission from bright, compact regions may trace photo-ionized gas in the immediate surroundings of massive stars. We take into consideration the point-like sources detected at 93~GHz as candidate massive star clusters, though some contamination by synchrotron-dominated supernova remnants or dusty protoclusters may still be possible.  The morphology of the 93~GHz emission and clustering of the peaks indicate possible ridges of star formation and shells.  The extended, faint negative bowls flanking the main disk likely reflect the short spacing data missing from this image. We do not expect that they affect our analysis of the point source-like cluster candidates. 

\begin{figure*}[t]
    \centering
    \includegraphics[width=0.49\textwidth]{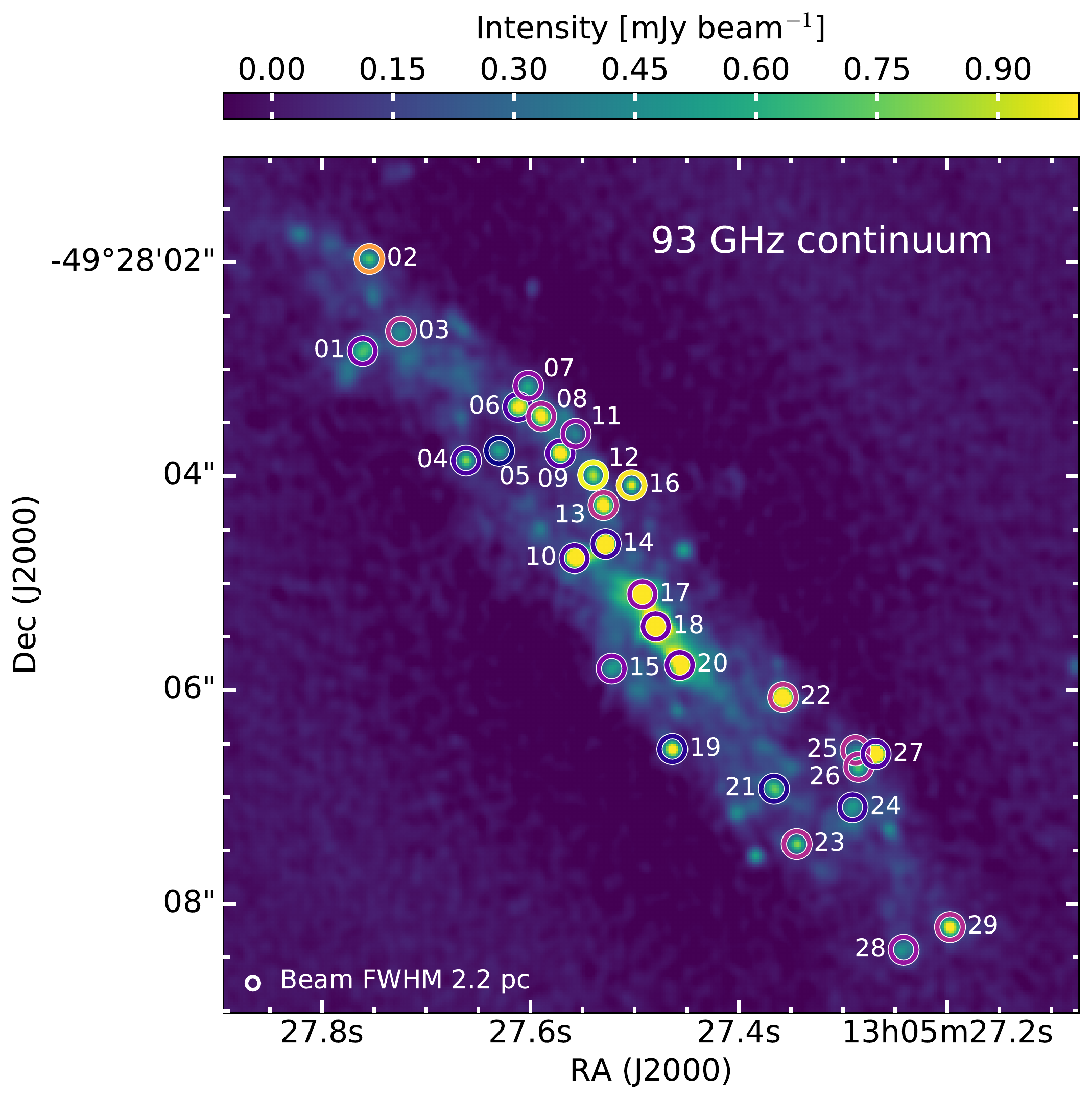}
    \includegraphics[width=0.49\textwidth]{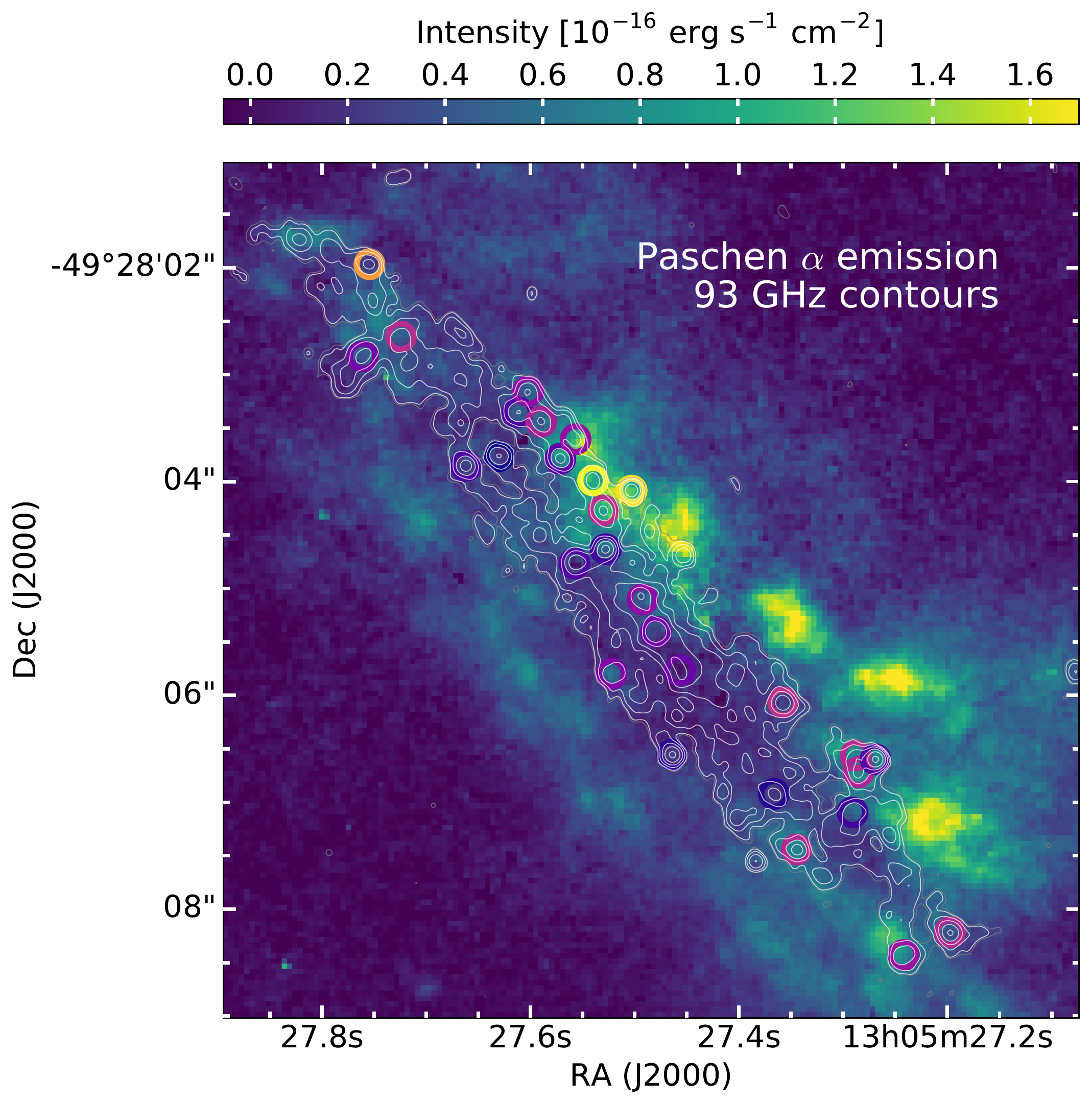} 
    \includegraphics[width=0.49\textwidth]{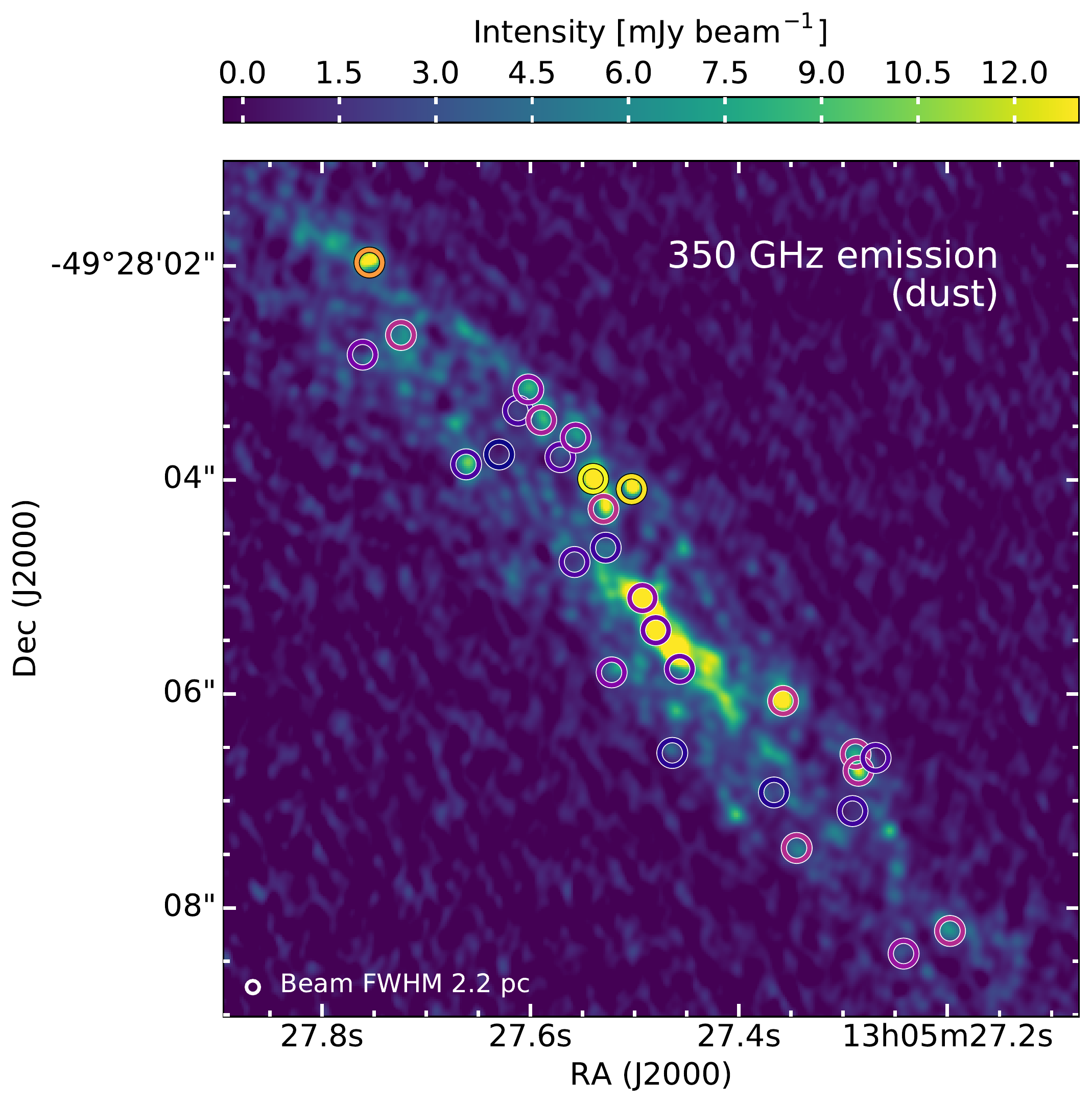}
    \includegraphics[width=0.49\textwidth]{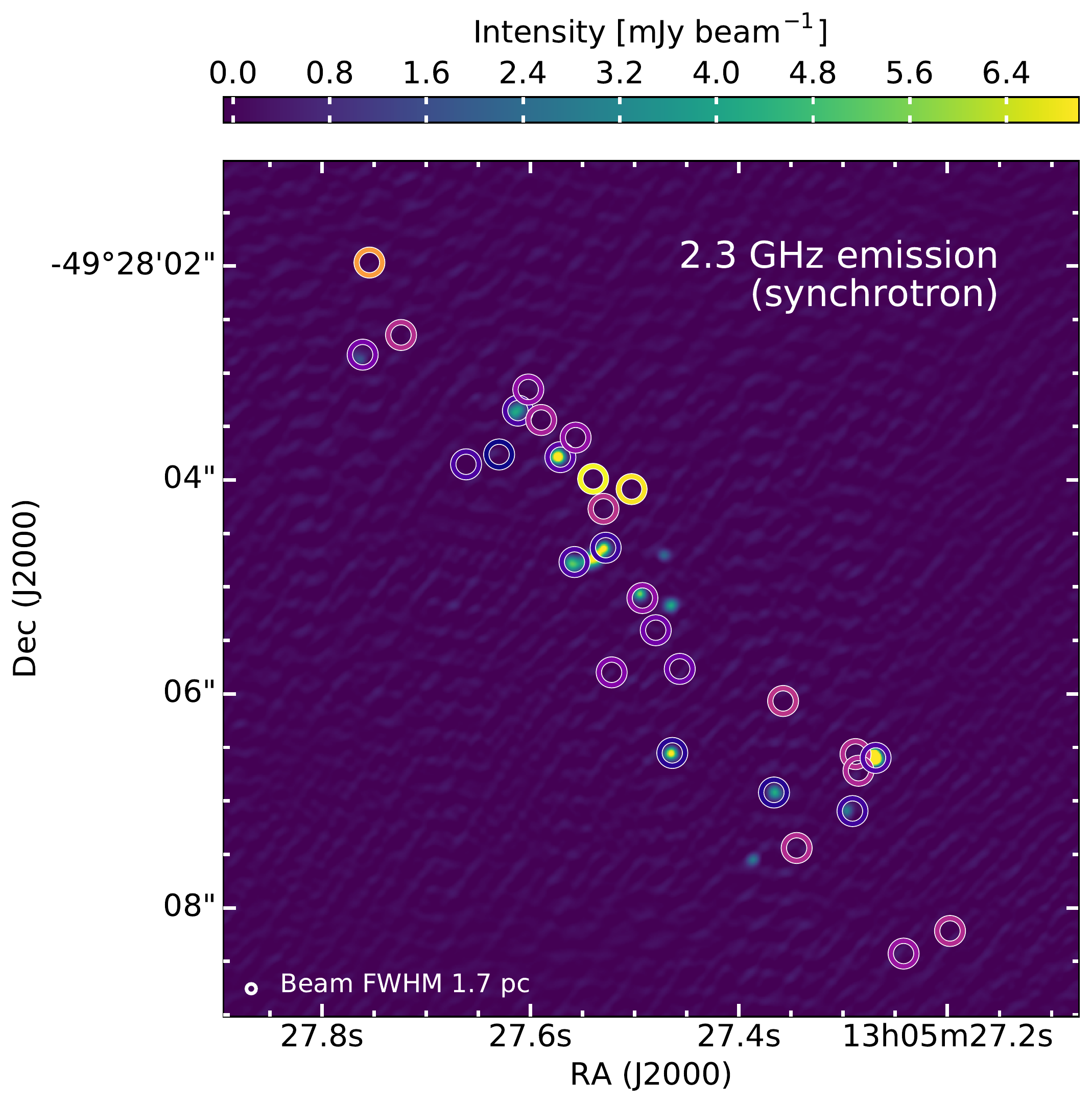}
    \caption{ \textit{Top left}: 93~GHz continuum emission with sources identified, also see Table~\ref{tab:continuum}. Circles show apertures (diameter of 0.24\arcsec) used for continuum extraction. Their colors indicate the measured in-band spectral index, as in Figure~\ref{fig:fluxcomp} where dark purple indicates synchrotron dominated emission and yellow indicates dust dominated emission. \textit{Top right}: HST Paschen-$\alpha$ emission -- hydrogen recombination line, $\mathsf{n} = 3$, at 1.87 $\mu$m -- (courtesy P. van der Werf) tracing ionized gas at $\approx 0.2$\arcsec\ resolution \citep{Marconi2000}. Dust extinction, of $A_{\mathrm{V}} > 36$~mag, obscures the Pa-$\alpha$ recombination emission at shorter wavelengths from the starburst region. Contours trace 93~GHz continuum, as described in Figure~\ref{fig:cont93}. \textit{Bottom left}: ALMA 350~GHz continuum emission tracing dust. \textit{Bottom right}: Australian LBA 2.3~GHz continuum imaging of synchrotron emission primarily from supernova remnants \citep{Lenc2009}. }
    \label{fig:multifreq}
\end{figure*}

The large amount of extinction present in this high inclination  central region \citep[ $i \sim 72^{\circ}$;][]{Henkel2018} has previously impeded the direct observation of its star clusters. Paschen-$\alpha$ (Pa-$\alpha$) emission \citep{Marconi2000} of the $\mathsf{n} = 3$ hydrogen recombination line at 1.87~$\mu$m, shown in Figure~\ref{fig:multifreq}, reveals faint emission above and below the star-forming plane. Corrected for extinction, the clumps of ionized emission traced by Pa-$\alpha$ would give rise to free-free emission 
below our ALMA detection limit. Pa-$\alpha$ along with mid-infrared spectral lines 
give support for dust extinction of $A_{\mathrm{V}} > 160$~mag surrounding the AGN core and more generally $A_{\mathrm{V}} \gtrsim 36$~mag in the star-forming region \citep{Spoon2000}.

A large fraction (18/29) of the 93~GHz sources coincide with peaks in dust emission at 350~GHz, as shown in Figure~\ref{fig:multifreq}. Overall there is a good correspondence between the two tracers. This indicates that candidate clusters are relatively young and may still harbor reservoirs of gas, though in Section~\ref{sec:totalmass} we find that the fraction of the mass still in gas tends to be relatively small.  

In Figure~\ref{fig:multifreq}, the 93~GHz peaks without dust counterparts tend to be strong sources of emission at 2.3 GHz \citep{Lenc2009}, a frequency where synchrotron emission typically dominates. As discussed in \cite{Lenc2009}, the sources at this frequency are predominantly supernova remnants. The presence of 13 possible supernova remnants -- four of which are resolved into shell-like structures with 1.1 to 2.1 pc in diameter -- indicates that a burst of star-formation activity started at least a few Myr ago. \cite{Lenc2009} modeled SEDs of the sources spanning 2.3--23 GHz and found significant opacity at 2.3 GHz ($\tau = 5- 22$), implying the presence of dense, free-free plasma in the vicinity of the supernova remnants. 

% AGN core
At 93~GHz, the very center of the starburst shows an elongated region of enhanced emission (about 20~pc in projected length, or $\sim$1\arcsec) that is also bright in 350~GHz emission. This region is connected to the areas of highest extinction. Higher column densities of ionized plasma are also present in the region; \cite{Lenc2009} observations reveal large free-free opacities at least up to 23~GHz. The brightest peak at 93~GHz, centered at $(\alpha, \delta)_{93} = (13\mathrm{h}\, 05\mathrm{m}\, 27.4798 \mathrm{s} \pm 0.004 \mathrm{s}, -49^{\circ}\, 28\arcmin\, 05.404\arcsec \pm 0.06\arcsec)$, is co-located with the kinematic center as determined from H$_{2}$O maser observations $(\alpha, \delta)_{\mathrm{H_{2}O}} = (13\mathrm{h}\, 05\mathrm{m}\, 27.279\mathrm{s} \pm 0.02\mathrm{s}, -49^{\circ}\, 28\arcmin\, 04.44\arcsec \pm 0.1\arcsec)$ \citep{Greenhill1997} and presumably harbors the AGN core.  We refer to the elongated region of enhanced emission surrounding the AGN core as the circumnuclear disk. The morphological similarities between 93~GHz and 350~GHz, together with the detection of a synchrotron point source (likely supernova remnant; see Section~\ref{sec:fluxextract} and Source 17) in the circumnuclear disk, indicate star-formation is likely present there.

%%%%%%%%%%%%%%%%%%%%%%%%%%%%%%%%%%%%%%%%
\subsection{Point Source Identification} 
\label{sec:ps_id}

We identify candidate star clusters via point-like sources \edit1{of emission} in the 93~GHz continuum image. Sources are found using \texttt{PyBDSF} \citep{Mohan2015} in the following way. Islands are defined as contiguous pixels \edit1{(of nine pixels or more)} above a threshold of seven times the global rms value of $\sigma \approx 0.017$~\mJyb. Within each island, multiple Gaussians may be fit, each with a peak amplitude greater than the peak threshold, a threshold of ten times the global rms. We chose this peak threshold to ensure that significant emission can also be identified in the continuum images made from individual spectral windows. The number of Gaussians is determined from the number of distinct peaks of emission higher than the peak threshold and which have a negative gradient in all eight evaluated directions. Starting with the brightest peak, Gaussians are fit and cleaned (i.e., subtracted). A source is identified with a Gaussian as long as subtracting its fit does not increase the island rms. 

Applying this algorithm to our 93~GHz image yielded $50$ Gaussian sources. We remove $5$ sources that fall outside of the star-forming region. We also removed $3$ sources that appeared blended, with an offset $<0.12$\arcsec\ from another source. Finally we remove $13$ sources that do not have a flux density above ten times the global rms after extracting the 93~GHz continuum flux density through aperture photometry (see Section~\ref{sec:fluxextract}). As a result, we analyze 29 sources as candidate star clusters. In Figure~\ref{fig:multifreq}, we show the location of each source with the apertures used for flux extraction. The sources match well with what we would identify by eye.

%%%%%%%%%%%%%%%%%%%%%%%%%%%%%%%%%%%%%%%%

\subsection{Point Source Flux Extraction} 
\label{sec:fluxextract}

For each source, we extract the continuum flux density at 2.3~GHz, 93~GHz, and 350~GHz through aperture photometry. Before extracting the continuum flux at 2.3~GHz, we convolve the image to the common resolution of 0.12\arcsec. We extract the flux density at the location of the peak source within an aperture diameter of 0.24\arcsec. Then we subtract the extended background continuum that is local to the source by taking the median flux density within an annulus of inner diameter 0.24\arcsec\ and outer diameter 0.30\arcsec; using the median suppresses the influence of nearby peaks and the bright surrounding filamentary features. The flux density of each source at each frequency is listed in Table~\ref{tab:continuum}. When the extracted flux density within an aperture is less than three times the global rms noise \edit1{(in the 2.3 GHz and 350 GHz images)}, we assign a three sigma upper limit to that flux measurement. 

\begin{figure}
    \centering
    \includegraphics[width=0.47\textwidth]{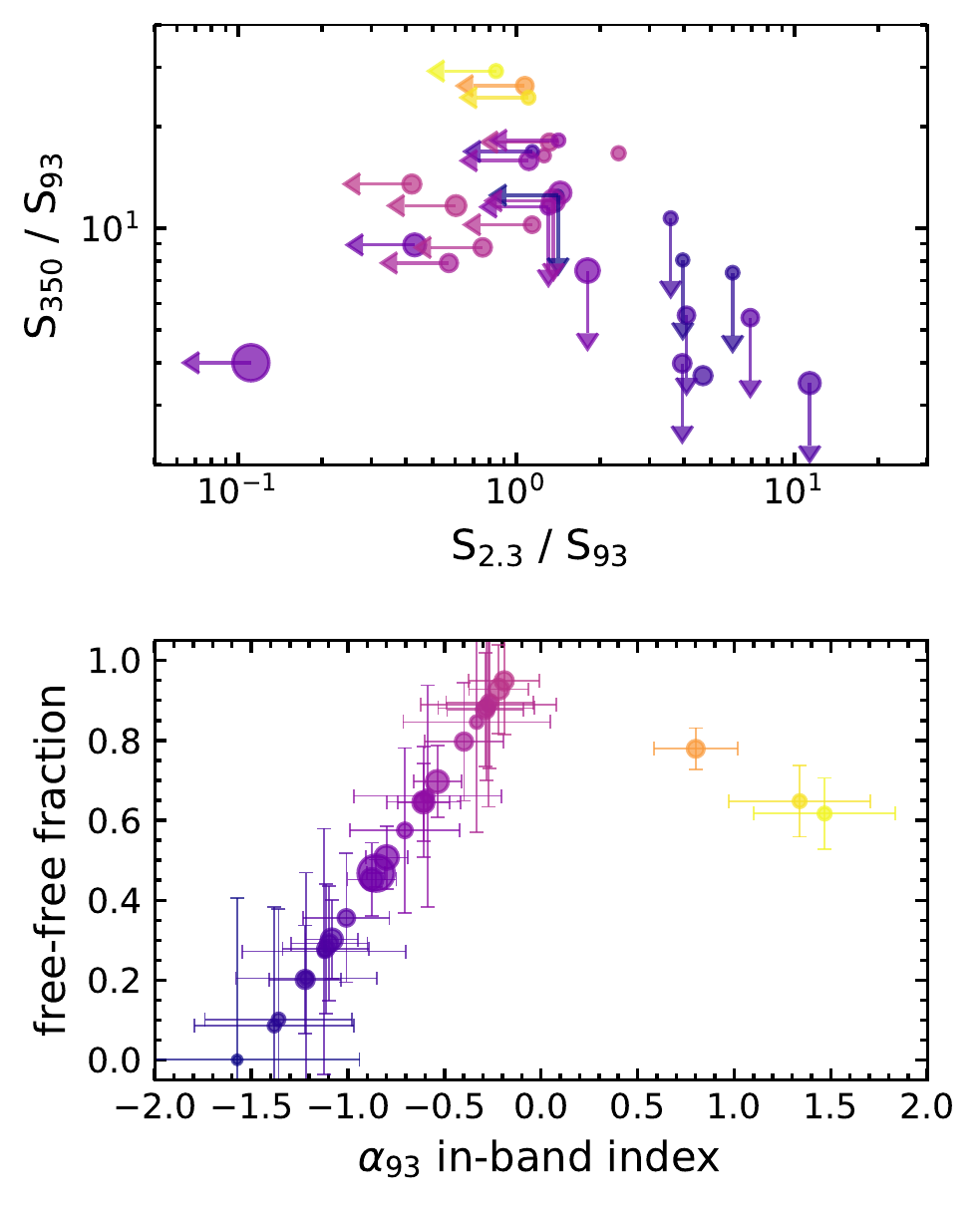}
    \caption{\textit{Top:} The ratio of the flux densities extracted at 350~GHz and 93~GHz (S$_{350}$/S$_{93}$) plotted against the ratio of the flux densities extracted at 2.3~GHz and 93~GHz (S$_{2.3}$/S$_{93}$), and \textit{bottom:} the relation we use to determine the free-free fraction from the in-band index, $\alpha_{93}$, at 93~GHz. Data points in both plots are colored by the in-band index derived only from our ALMA data. Yellow indicates dust dominated sources, whereas purple indicates synchrotron dominated sources. Candidate star clusters in which free-free emission dominates at 93~GHz appear $\sim$pink. The diameter of each data point is proportional to the flux density at 93~GHz and correspondingly inversely proportional to the error of the in-band spectral index.}
    \label{fig:fluxcomp}
\end{figure}

In Figure~\ref{fig:fluxcomp}, we plot the ratio of the flux densities extracted at 350~GHz and 93~GHz (S$_{350}$/S$_{93}$) against the ratio of the flux densities extracted at 2.3~GHz and 93~GHz (S$_{2.3}$/S$_{93}$). Synchrotron dominated sources, which fall to the bottom right of the plot, separate from the free-free (and dust) dominated sources, which lie in the middle of the plot. One exception is the AGN (Source 18) which, due to self-absorption at frequencies greater than 23~GHz, is bright at 93~GHz but not at 2.3~GHz and therefore has the lowest S$_{2.3}$/S$_{93}$ ratio. 

\begin{figure}[t]
    \includegraphics[width=0.47\textwidth]{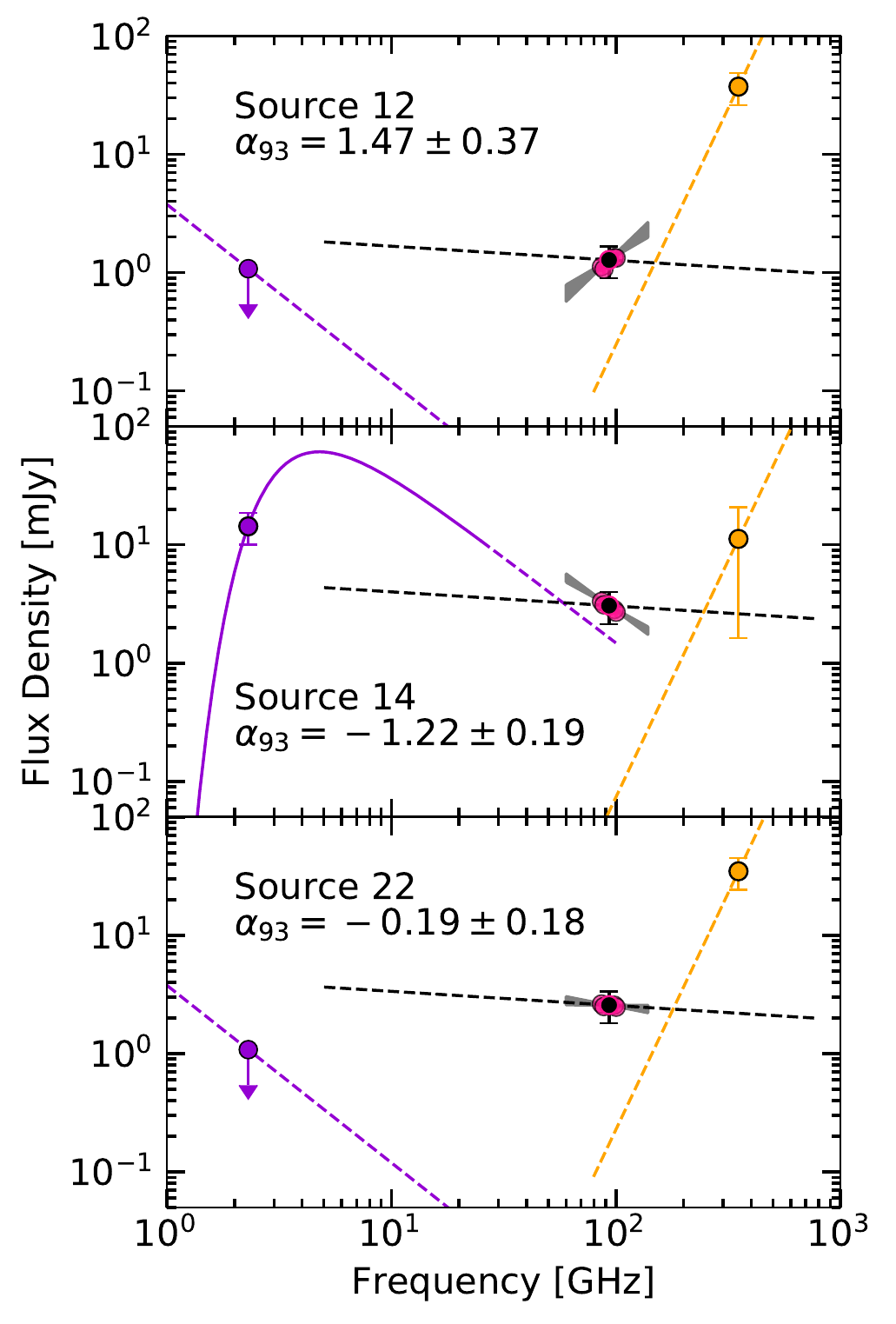}
    \caption{Example SEDs constructed for each source. \textit{Top:} Dust dominated, Source 12. \textit{Middle:} Synchrotron dominated, Source 14. \textit{Bottom:} Free-free dominated, Source 22. The dashed orange line represents a dust spectral index of $\alpha = 4.0$, normalized to the flux density we extract at 350~GHz (orange data point). The dashed black line represents a free-free spectral index of $\alpha = -0.12$,  normalized to the flux density we extract at 93~GHz (black data point). The \edit1{pink} data points show the flux densities extracted from the band 3 spectral windows. The gray shaded region is the 1$\sigma$ error range of the band 3 spectral index fit, except we have extended the fit in frequency for displaying purposes. The purple line represents a synchrotron spectral index of $\alpha = -1.5$,  normalized to the flux density we extract at 2.3~GHz (purple data point); except for Source 14 where the solid purple line represents the normalized 2.3--23 GHz fit from \cite{Lenc2009}.  Error bars on the flux density data points are 3$\sigma$. }
    \label{fig:SED}
\end{figure}

From the continuum measurements we construct simple SEDs for each source. \edit1{These SEDs are used for illustrative purposes and do not affect the analysis in this paper.} Examples of the SEDs of three sources are included in Figure~\ref{fig:SED}. We show an example of a free-free dominated source (Source 22), which represents the majority of sources, as well as a dust (Source 12) and a synchrotron (Source 14) dominated source. Of the sources \edit1{with extracted emission of $>$3$\sigma$ at 2.3~GHz,} nine also have the free-free absorption of their synchrotron spectrum modeled. We \edit1{plot} this information whenever possible. When a source identified by \cite{Lenc2009} lies within 0.06\arcsec\ (half the beam FWHM) of the 93~GHz source, we associate the low-frequency modeling with the 93~GHz source. We take the model fit by \cite{Lenc2009} and normalize it to the 2.3~GHz flux that we extract \edit1{--} as an example, see the solid purple curve in the middle panel of Figure~\ref{fig:SED}. The SEDs of all sources are shown in Figure~\ref{fig:ap_sed} in Appendix~\ref{ap:sed}.

\begin{deluxetable*}{l r r R R R R R R R}
    \setlength{\tabcolsep}{5pt}
    \tablecaption{Properties of the continuum emission from candidate star clusters. \label{tab:continuum}}
    \tablewidth{0pt}
    \tablehead{
    \colhead{Source} %1
    & \colhead{RA} %2
    & \colhead{Dec} %3
    & \colhead{$S_{93}$} %4
    & \colhead{$\alpha_{93}$} %5
    & \colhead{$S_{2.3}$~$^{a}$} %6
    & \colhead{$S_{350}$~$^{b}$} %7
    & \colhead{$f_{\mathrm{ff}}$} %8
    & \colhead{$f_{\mathrm{syn}}$~$^c$} %9
    & \colhead{$f_{\mathrm{d}}$~$^c$} \\ %10
    \colhead{} & \colhead{} & \colhead{} & \colhead{(mJy)} & \colhead{} & \colhead{(mJy)} & \colhead{(mJy)} & \colhead{} & \colhead{} & \colhead{}
    }
    \startdata
01 & 13:05:27.761 & $-$49:28:02.83 & 1.28 \pm 0.13 & -0.80 \pm 0.11 & 2.3 \pm 0.4 & ... & 0.51 \pm 0.08 & 0.49 & ...  \\ 
02 & 13:05:27.755 & $-$49:28:01.97 & 1.01 \pm 0.10 & 0.80 \pm 0.22 & ... & 26.6 & 0.78 \pm 0.05 & ... & 0.22  \\ 
03 & 13:05:27.724 & $-$49:28:02.64 & 0.82 \pm 0.08 & -0.27 \pm 0.23 & ... & 14.8 & 0.89 \pm 0.16 & 0.11 & ...  \\ 
04 & 13:05:27.662 & $-$49:28:03.85 & 0.95 \pm 0.09 & -1.12 \pm 0.42 & ... & 16.0 & 0.27 \pm 0.31 & 0.73 & ...  \\ 
05 & 13:05:27.630 & $-$49:28:03.76 & 0.77 \pm 0.08 & -1.57 \pm 0.63 & ... & ... & 0.00 \pm 0.40 & 1.00 & ...  \\ 
06 & 13:05:27.612 & $-$49:28:03.35 & 1.73 \pm 0.17 & -1.11 \pm 0.22 & 7.1 \pm 0.7 & ... & 0.28 \pm 0.16 & 0.72 & ...  \\ 
07 & 13:05:27.602 & $-$49:28:03.15 & 0.98 \pm 0.10 & -0.61 \pm 0.19 & ... & 15.5 & 0.65 \pm 0.14 & 0.35 & ...  \\ 
08 & 13:05:27.590 & $-$49:28:03.44 & 1.89 \pm 0.19 & -0.40 \pm 0.20 & ... & 14.9 & 0.80 \pm 0.15 & 0.20 & ...  \\ 
09 & 13:05:27.571 & $-$49:28:03.78 & 1.76 \pm 0.18 & -1.01 \pm 0.22 & 12.2 \pm 1.2 & ... & 0.36 \pm 0.16 & 0.64 & ...  \\ 
10 & 13:05:27.558 & $-$49:28:04.77 & 2.41 \pm 0.24 & -1.10 \pm 0.20 & 9.5 \pm 0.9 & ... & 0.29 \pm 0.14 & 0.71 & ...  \\ 
11 & 13:05:27.557 & $-$49:28:03.60 & 0.76 \pm 0.08 & -0.59 \pm 0.38 & ... & 13.9 & 0.66 \pm 0.28 & 0.34 & ...  \\ 
12 & 13:05:27.540 & $-$49:28:03.99 & 1.28 \pm 0.13 & 1.47 \pm 0.37 & ... & 37.4 & 0.62 \pm 0.09 & ... & 0.38  \\ 
13 & 13:05:27.530 & $-$49:28:04.27 & 1.78 \pm 0.18 & -0.22 \pm 0.15 & ... & 20.8 & 0.93 \pm 0.11 & 0.07 & ...  \\ 
14 & 13:05:27.528 & $-$49:28:04.63 & 3.06 \pm 0.31 & -1.22 \pm 0.19 & 14.3 \pm 1.4 & 11.2 & 0.20 \pm 0.14 & 0.80 & ...  \\ 
15 & 13:05:27.522 & $-$49:28:05.80 & 0.83 \pm 0.08 & -0.71 \pm 0.28 & ... & ... & 0.57 \pm 0.21 & 0.42 & ...  \\ 
16 & 13:05:27.503 & $-$49:28:04.08 & 0.98 \pm 0.10 & 1.34 \pm 0.37 & ... & 23.9 & 0.65 \pm 0.09 & ... & 0.35  \\ 
17 & 13:05:27.493 & $-$49:28:05.10 & 3.69 \pm 0.37 & -0.61 \pm 0.13 & 5.3 \pm 0.5 & 47.1 & 0.65 \pm 0.10 & 0.35 & ...  \\ 
18 & 13:05:27.480 & $-$49:28:05.40 & 9.74 \pm 0.97 & -0.85 \pm 0.05 & ... & 39.0 & 0.47 \pm 0.03 & 0.53 & ...  \\ 
19 & 13:05:27.464 & $-$49:28:06.55 & 1.30 \pm 0.13 & -1.36 \pm 0.38 & 7.8 \pm 0.8 & ... & 0.10 \pm 0.28 & 0.90 & ...  \\ 
20 & 13:05:27.457 & $-$49:28:05.76 & 2.51 \pm 0.25 & -0.88 \pm 0.13 & ... & 22.4 & 0.45 \pm 0.09 & 0.55 & ...  \\ 
21 & 13:05:27.366 & $-$49:28:06.92 & 1.19 \pm 0.12 & -1.38 \pm 0.41 & 4.7 \pm 0.5 & ... & 0.09 \pm 0.30 & 0.91 & ...  \\ 
22 & 13:05:27.358 & $-$49:28:06.07 & 2.57 \pm 0.26 & -0.19 \pm 0.18 & ... & 34.8 & 0.95 \pm 0.13 & 0.05 & ...  \\ 
23 & 13:05:27.345 & $-$49:28:07.44 & 0.95 \pm 0.09 & -0.28 \pm 0.25 & ... & 9.7 & 0.88 \pm 0.18 & 0.12 & ...  \\ 
24 & 13:05:27.291 & $-$49:28:07.10 & 0.90 \pm 0.09 & -1.22 \pm 0.36 & 3.2 \pm 0.4 & ... & 0.20 \pm 0.26 & 0.80 & ...  \\ 
25 & 13:05:27.288 & $-$49:28:06.56 & 0.91 \pm 0.09 & -0.27 \pm 0.35 & 2.1 \pm 0.4 & 15.2 & 0.89 \pm 0.26 & 0.11 & ...  \\ 
26 & 13:05:27.285 & $-$49:28:06.72 & 1.13 \pm 0.11 & -0.33 \pm 0.38 & 1.4 \pm 0.4 & 18.5 & 0.85 \pm 0.28 & 0.15 & ...  \\ 
27 & 13:05:27.269 & $-$49:28:06.60 & 2.75 \pm 0.28 & -1.08 \pm 0.14 & 31.2 \pm 3.1 & ... & 0.30 \pm 0.10 & 0.70 & ...  \\ 
28 & 13:05:27.242 & $-$49:28:08.43 & 0.80 \pm 0.08 & -0.54 \pm 0.12 & ... & ... & 0.70 \pm 0.09 & 0.30 & ...  \\ 
29 & 13:05:27.198 & $-$49:28:08.22 & 1.43 \pm 0.14 & -0.29 \pm 0.20 & ... & 12.6 & 0.88 \pm 0.14 & 0.12 & ...  \\ 
\enddata
\tablecomments{RA and Dec refer to the center location of a Gaussian source identified in the 93~GHz continuum image, in units of hour angle and degrees, respectively. $S_{93}$ is the flux density in the 93~GHz full bandwidth continuum image. $\alpha_{93}$ is the spectral index at 93~GHz ($S\propto \nu^{\alpha}$), as determined from the best fit slope to the 85--101~GHz continuum emission. $S_{2.3}$ is the flux density extracted in the 2.3~GHz continuum image. $S_{350}$ is the flux density extracted in the 350~GHz continuum image. $f_{\mathrm{ff}}$, $f_{\mathrm{syn}}$, $f_{\mathrm{d}}$ are the free-free, synchrotron, and dust fractional contribution to the 93~GHz continuum emission, respectively, as determined from the spectral index. See Section~\ref{sec:freefreefrac} for a description of how the errors on these fractional estimates are determined. \\
$^a$ A 3$\sigma$ upper limit to the sources undetected in 2.3~GHz continuum emission is 1.2~mJy. \\
$^b$ The error on the 350~GHz flux density measurement is 3.2~mJy. A 3$\sigma$ upper limit to the undetected sources is 9.6~mJy. \\
$^c$ The error on the fractional contributions are the same as for $f_{\mathrm{ff}}$ unless otherwise noted.
}
\end{deluxetable*}

%%%%%%%%%%%%%%%%%%%%%%%%%%%%%%%%%%%%%%%%

\subsection{Free-free Fraction at 93~GHz}
\label{sec:freefreefrac}

The flux density of optically thin, free-free emission at millimeter wavelengths \citep[see Appendix~\ref{ap:cont};][]{Draine2011} arises as
\begin{equation}
    \begin{split}
    S_{\mathrm{ff}} = & (2.08~\mathrm{mJy} ) 
	\left( \frac{ n_e n_+ V }{ 5\times10^8~\mathrm{cm^{-6}~pc^3} } \right) \times \\
	& \left(  \frac{T_e}{ 10^4~\mathrm{K}} \right)^{-0.32} 
	\left( \frac{\nu}{ 100~\mathrm{GHz} } \right)^{-0.12}
	\left( \frac{ D }{ 3.8~\mathrm{Mpc} } \right)^{-2}
    \label{eq:Scont}
    \end{split}
\end{equation}
where $EM_{\mathrm{C}} = n_e n_+ V$ is the volumetric emission measure of the ionized gas, $D$ is the distance to the source, and $T_e$ is the electron temperature of the medium. Properties of massive star clusters (i.e., ionizing photon rate) can thus be derived through an accurate measurement of the free-free flux density and an inference of the volumetric emission measure.  We will determine a free-free fraction, $f_{\mathrm{ff}}$, and let $S_{\mathrm{ff}} = f_{\mathrm{ff}} S_{93}$. 

In this section we focus on determining the portion of free-free emission that is present in the candidate stars clusters at 93~GHz. To do this, we need to estimate and remove contributions from synchrotron emission and dust continuum. We determine an in-band spectral index across the 15~GHz bandwidth of the ALMA Band 3 observations. Using the spectral index\footnote{\edit1{similar methods have been used by e.g., \cite{Linden2020}}}, we constrain the free-free fraction as well as the fractional contributions of synchrotron and dust (see Figure~\ref{fig:fluxcomp} and Table~\ref{tab:continuum}). We found the in-band index \edit1{to give stronger constraints and, for 2.3 GHz, to be more reliable than extrapolating (assuming indices of $-0.8$,$-1.5$)} because of the two decades difference in frequency coupled with the large optical depths already present at 2.3~GHz.

\subsubsection{Band 3 Spectral Index}

We estimate an in-band spectral index at 93~GHz, $\alpha_{93}$, from a fit to the flux densities in the spectral window continuum images of the ALMA Band 3 data. 
The spectral windows span 15 GHz, which we set up to have a large fractional bandwidth. We extract the continuum flux density of each source in each spectral window using aperture photometry with the same aperture sizes as described in Section~\ref{sec:fluxextract}. We fit a first order polynomial to the five continuum measurements -- four from each spectral window, one from the full bandwidth image. The fit to the in-band spectral index is listed in Table~\ref{tab:continuum} with the one sigma uncertainty to the fit. The median uncertainty of the spectral indices is 0.13.  The spectral indices measured in this way were consistent for the brightest sources with the index determined by \texttt{CASA tclean}; however, we found our method to be more reliable for the fainter sources. Nonetheless, the errors are large for faint sources. 

\subsubsection{Decomposing the fractional contributions of emission type}

From the in-band spectral index fit, we estimate the fractional contribution of each emission mechanism --- free-free, synchrotron, and (thermal) dust --- to the 93~GHz continuum. 
To do this, we simulate how mixtures of synchrotron, free-free, and dust emission could combine to create the observed in-band index. \edit2{ We assume fixed spectral indices for each component and adjust their fractions to reproduce the observations (see below).}

\edit1{ We consider that a source is dominated by two types of emission (a caveat which we discuss in detail in Section~\ref{sec:discuss}): free-free and dust, or free-free and synchrotron.
With synthetic data points, we first set the flux density at 93~GHz, $S_{93}$, \edit2{to a fixed value} and vary the contributions of dust and free-free continua, such that $S_{93} = S_{d} + S_{ff}$. For each model, we determine the continuum flux at each frequency across the Band 3 frequency coverage as the sum of the two components. We assume that the frequency dependence of the free-free component is $\alpha_{\mathrm{ff}} = -0.12$ and the frequency dependence of the dust component is $\alpha_{\mathrm{d}} = 4.0$. We fit the (noise-less) continuum of the synthetic data across the Band 3 frequency coverage with a power-law, determining the slope as $\alpha_{93}$, the in-band index. We express the results in terms of a free-free fraction and dust fraction, rather than absolute flux. \edit2{In this respect, we explore free-free fractions to the 93~GHz continuum ranging from $f_{\mathrm{ff}} = (0.001,0.999)$ in steps of $0.001$.} We let the results of this process constrain our free-free fraction when the in-band index measured in the actual (observed) data is $\alpha_{93} \geq -0.12$. }

\edit1{Next, we repeat the exercise, but we let synchrotron and free-free dominate the contribution to the continuum at 93~GHz. We take the frequency dependence of the free-free component as $\alpha_{\mathrm{ff}} = -0.12$ across the Band 3 frequency coverage, and we set the synchrotron component to $\alpha_{\mathrm{syn}} = -1.5$. We let the results of this process constrain the free-free fraction of the candidate star clusters when the fit to their in-band index is $\alpha_{93} \leq -0.12$.}

\edit1{ Our choices for the spectral indices of the three emission types are motivated as follows. In letting, $\alpha_{\mathrm{ff}} = -0.12$ we assume that the free-free emission is optically thin (e.g., see Appendix~\ref{ap:cont}). We do not expect significant free-free opacity at 93~GHz given the somewhat-evolved age of the candidate clusters in the starburst (see Section~\ref{sec:age}). In letting $\alpha_{\mathrm{d}} = 4.0$, we assume the dust emission is optically thin with a wavelength-dependent emissivity so that $\tau \propto \lambda^{-2}$ \citep[e.g., see][]{Draine2011}. The low optical dust optical depths estimated in Section~\ref{sec:dustgas}
imply a dust spectral index steeper than $2$, though the exact value might not be $4$, e.g., if our assumed emissivity power law index is not applicable. The generally faint dust emission indicates that our assumptions about dust do not have a large effect on our results. For the synchrotron frequency dependence, we assume $\alpha_{\mathrm{syn}} = -1.5$. This value is consistent with the best fit slope of $\alpha_{\mathrm{syn}} = -1.4$ found by \cite{Bendo2016} averaged over the central 30\arcsec\ of \trg. Furthermore, the median slope of synchrotron-dominated sources modeled at 2.3--23 GHz is -1.11 \citep{Lenc2009}, indicating that even at 23 GHz the synchrotron spectra already show losses due to aging, i.e., are steeper than a canonical initial injection of $\alpha_{\mathrm{syn}} \approx -0.8$. While our assumed value of $\alpha_{\mathrm{syn}} = -1.5$ is well-motivated on average, variations from source to source are likely present. }

\edit1{
Through this method of decomposition, the approximate relation between the in-band index and the free-free fraction is
\begin{equation}
    f_{\mathrm{ff}} = 
\begin{dcases}
    0.72\,\alpha_{93} + 1.09 , & -1.5 \leq \alpha_{93} \leq -0.12 \\
   -0.24\,\alpha_{93} + 0.97 , & 4.0 \geq \alpha_{93} \geq -0.12 \\
\end{dcases}
\end{equation}
for which $\alpha_{93} \leq -0.12$, the synchrotron fraction is found to be $f_{\mathrm{syn}} = 1 - f_{\mathrm{ff}}$ and we set $f_{\mathrm{d}} = 0$, and for which $\alpha_{93} \geq -0.12$ the dust fraction is found to be $f_{\mathrm{d}} = 1 - f_{\mathrm{ff}}$ and $f_{\mathrm{syn}} = 0$.
This relation is depicted in the bottom panel of Figure~\ref{fig:fluxcomp} using the values that have been determined for each source.}

\subsubsection{Estimated free-free, synchrotron, and dust fractions to the 93~GHz continuum}

Table~\ref{tab:continuum} and Figure~\ref{fig:fluxcomp} summarize our estimated fractional contribution of each emission mechanism to the 93~GHz continuum of each source. \edit1{We use the same relation above to translate the range of uncertainty on the spectral index to an uncertainty in the emission fraction estimates.} We find, at this 0.12\arcsec\ resolution, the median free-free fraction of sources is $f_{\mathrm{ff}} = 0.62$ \edit1{with median absolute deviation of 0.29}. Most of the spectral indices are negative, and as a result, we find synchrotron emission can have a non-trivial contribution with a median fraction of $f_{\mathrm{syn}} = 0.36$ \edit1{and median absolute deviation of 0.32}.  On the other hand, three of the measured spectral indices are positive. \edit1{The median dust fraction of sources is $f_{\mathrm{d}} = 0$ and median absolute deviation of 0.10. A} $1\sigma$ limit on the fractional contribution of dust does not exceed $f_{\mathrm{d}} = 0.47$ for any \edit1{single} source.

Additional continuum observations of comparable resolution at frequencies between 2~GHz and 350~GHz would improve the estimates of the fractional contribution of free-free, dust and synchrotron to the 93~GHz emission. 

%%%%%%%%%%%%%%%%%%%%%%%%%%%%%%%%%%%%%%%%

\section{Recombination Line Emission}   \label{sec:spectra}

Hydrogen recombination lines at these frequencies trace ionizing radiation ($E >13.6$~eV); this recombination line emission is unaffected by dust extinction. The integrated emission from a radio recombination line transition to quantum number $\mathsf{n}$, which we derive for millimeter wavelength transitions in Appendix~\ref{ap:line}, is described by 
\begin{equation}
    \begin{split}
    \int S_{{\mathsf{n}}}~\mathrm{dv} =  
	& \left( 65.13~\mathrm{mJy~km~s^{-1}}\right) \times \\
 	& b_{\mathsf{n+1}} 
	\left( \frac{n_e n_p V }{ 5\times10^8~\mathrm{cm^{-6}~pc^3} }\right)
	\left( \frac{ D }{ 3.8~\mathrm{Mpc} }\right)^{-2} \times \\
	& \left( \frac{ T_e }{ 10^4~\mathrm{K} } \right)^{-1.5} 
	\left( \frac{\nu}{100~\mathrm{GHz}} \right)
    \label{eq:Sline}
    \end{split}
\end{equation}
where $b_{\mathsf{n+1}}$ is the LTE departure coefficient,  $EM_{\mathrm{L}} = n_e n_p V$ is the volumetric emission measure of ionized hydrogen, $D$ is the distance to the source, $T_e$ is the electron temperature of the ionized gas, and $\nu$ is the rest frequency of the spectral line.

\begin{figure*}
    \centering
    \includegraphics[width=0.95\textwidth]{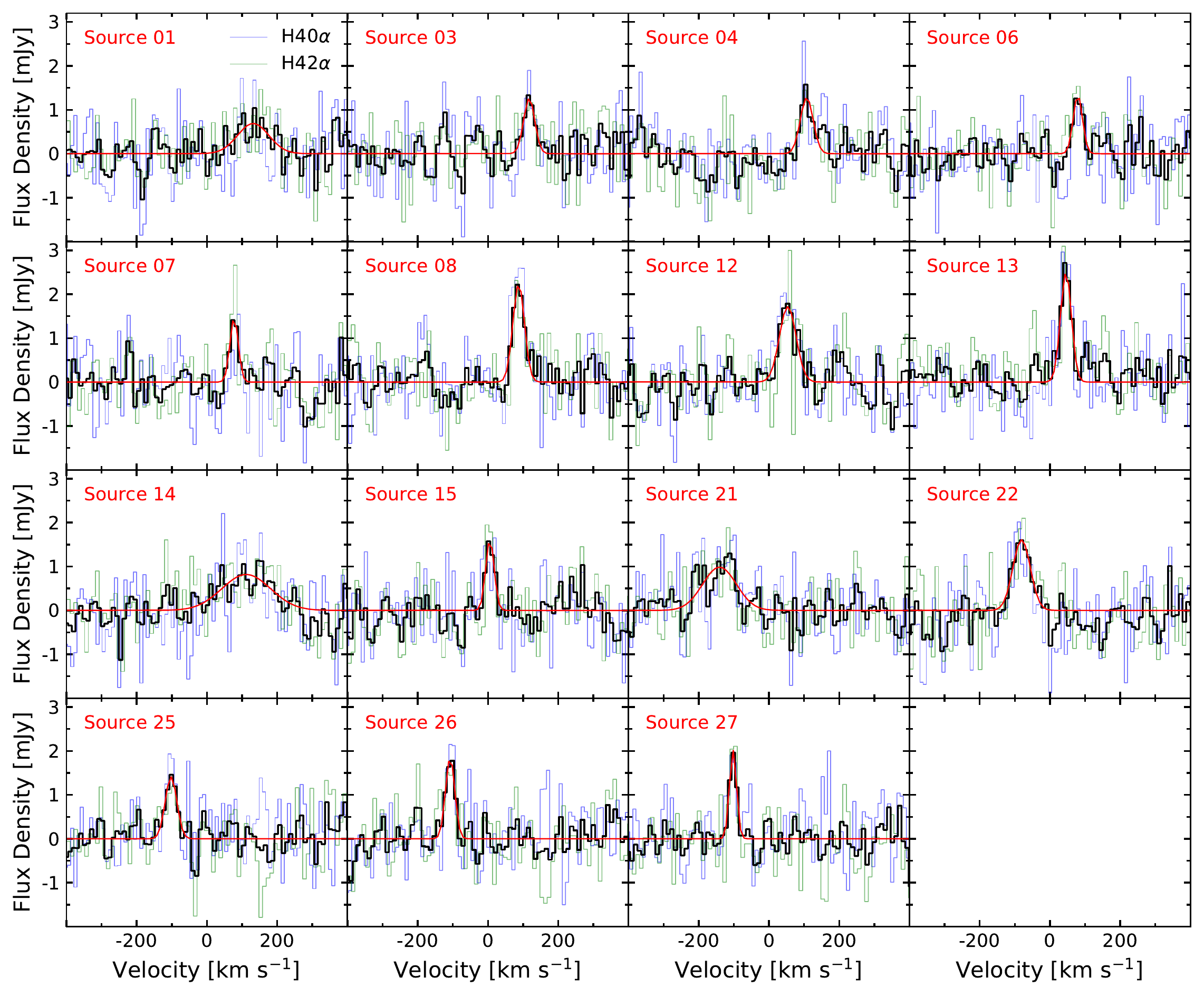}
    \caption{Radio recombination line spectra for sources with significantly detected emission. The thin blue line is the H40$\alpha$ spectrum. The thin green line is the H42$\alpha$. These spectra have been regridded from their native velocity resolution to the common resolution of 10.3~\kms. The thick black line is the weighted average spectrum of H40$\alpha$ and H42$\alpha$, effectively H41$\alpha$. In red is the best fit to the effective H41$\alpha$ radio recombination line feature.}
    \label{fig:spectra}
\end{figure*}

Figure~\ref{fig:spectra} shows spectra of the 15 sources with detected radio recombination line emission. We extract H40$\alpha$ and H42$\alpha$ spectra at the location of each source with an aperture diameter of 0.4\arcsec, or twice the beam FWHM. We average the spectra of the two transitions together to enhance the signal-to-noise ratio of the recombination line emission. To synthesize an effective H41$\alpha$ profile, we interpolate the two spectra of each source to a fixed velocity grid with a channel width of 10.3~\kms, weight each spectrum by $\sigma_{\mathrm{rms}}^{-2}$ where $\sigma_{\mathrm{rms}}$ is the spectrum standard deviation, and average the spectra together lowering the final noise. The averaged spectrum has an effective transition of H41$\alpha$ at $\nu_{\mathrm{eff}} = 92.034$~GHz. 

We fit spectral features with a Gaussian profile. We calculate an integrated signal-to-noise ratio for each line by integrating the spectrum across the Gaussian width of the fit (i.e., $\pm \sigma_{\mathrm{Gaus}}$) and then dividing by the noise over the same region, $\sqrt{N} \sigma_{\mathrm{rms}}$, where $N$ is the number of channels covered by the region. We report on detections with an integrated signal of $>5\sigma_{\mathrm{rms}}$. Table~\ref{tab:line} summarizes the properties of the line profiles derived from the best-fit Gaussian. The median rms of the spectra is $\sigma_{\mathrm{rms}} = 0.34$~mJy. 

\begin{deluxetable}{c C C C C}
    %\tablenum{1}
    \tablecaption{Average Line Profiles nominally located near H41$\alpha$ \label{tab:line}}
    \tablewidth{0pt}
    \tablehead{
    \colhead{Source} %1
    & \colhead{V$_{cen}$} %2
    & \colhead{Peak} %3
    & \colhead{FWHM} %4
    & \colhead{$\sigma_{\mathrm{rms}}$} \\ %5
    \colhead{} & \colhead{(\kms)} & \colhead{(mJy)} & \colhead{(\kms)} & \colhead{(mJy)}
    }
    %\decimalcolnumbers
    \startdata
01 & 131.6 \pm 12 & 0.69 \pm 0.16 & 105.2 \pm 28 & 0.35  \\ 
03 & 117.4 \pm 4.0 & 1.26 \pm 0.28 & 37.2 \pm 9.4 & 0.36  \\ 
04 & 107.5 \pm 4.3 & 1.28 \pm 0.26 & 42.9 \pm 10 & 0.36  \\ 
06 & 79.4 \pm 3.1 & 1.33 \pm 0.27 & 31.1 \pm 7.3 & 0.33  \\ 
07 & 77.2 \pm 3.0 & 1.46 \pm 0.32 & 28.1 \pm 7.0 & 0.36  \\ 
08 & 87.5 \pm 2.1 & 2.24 \pm 0.24 & 38.7 \pm 4.9 & 0.33  \\ 
12 & 53.6 \pm 3.9 & 1.72 \pm 0.24 & 55.9 \pm 9.1 & 0.39  \\ 
13 & 45.7 \pm 1.7 & 2.68 \pm 0.26 & 34.8 \pm 3.9 & 0.31  \\ 
14 & 111.8 \pm 12 & 0.82 \pm 0.12 & 163.1 \pm 28 & 0.33  \\ 
15 & 5.6 \pm 2.8 & 1.62 \pm 0.33 & 28.2 \pm 6.6 & 0.37  \\ 
21 & -140.7 \pm 8.3 & 0.99 \pm 0.15 & 113.3 \pm 20 & 0.34  \\ 
22 & -80.6 \pm 4.0 & 1.61 \pm 0.22 & 58.4 \pm 9.4 & 0.34  \\ 
25 & -102.3 \pm 3.3 & 1.42 \pm 0.24 & 39.0 \pm 7.7 & 0.32  \\ 
26 & -107.6 \pm 2.2 & 1.86 \pm 0.27 & 30.9 \pm 5.3 & 0.33  \\ 
27 & -102.3 \pm 1.6 & 2.02 \pm 0.27 & 23.6 \pm 3.6 & 0.28  \\ 
    \enddata
    \tablecomments{V$_{\mathrm{cen}}$ is the central velocity of the best fit Gaussian. Peak is the peak amplitude of the Gaussian fit. FWHM is the full-width half maximum of the Gaussian fit. $\sigma_{\mathrm{rms}}$ is the standard deviation of the fit-subtracted spectrum. }
\end{deluxetable}

In Figure~\ref{fig:ap_allspec} of Appendix~\ref{ap:spec}, we show that the central velocities of our detected recombination lines are in good agreement with the kinematic velocity expected of the disk rotation. To do this, we overlay our spectra on H40$\alpha$ spectra extracted from the intermediate configuration observations (0.7\arcsec\ resolution).

In \edit1{12 of the 15} sources, we detect relatively narrow features of $\mathrm{FWHM} \sim (24 - 58)$~\kms. Larger line-widths of $\mathrm{FWHM} \sim (105 - 163)$~\kms\ are observed from bright sources which also have high synchrotron fractions, indicating that multiple components, unresolved motions (e.g., from expanding shells or galactic rotation), or additional turbulence may be present. Six sources with detected recombination line emission have considerable ($f_{\mathrm{syn}} \gtrsim 0.50$) synchrotron emission (i.e. Sources 1, 4, 6, 14, 21, 27) at 93~GHz. 

\begin{figure}
    \centering
    \includegraphics[width=0.44\textwidth]{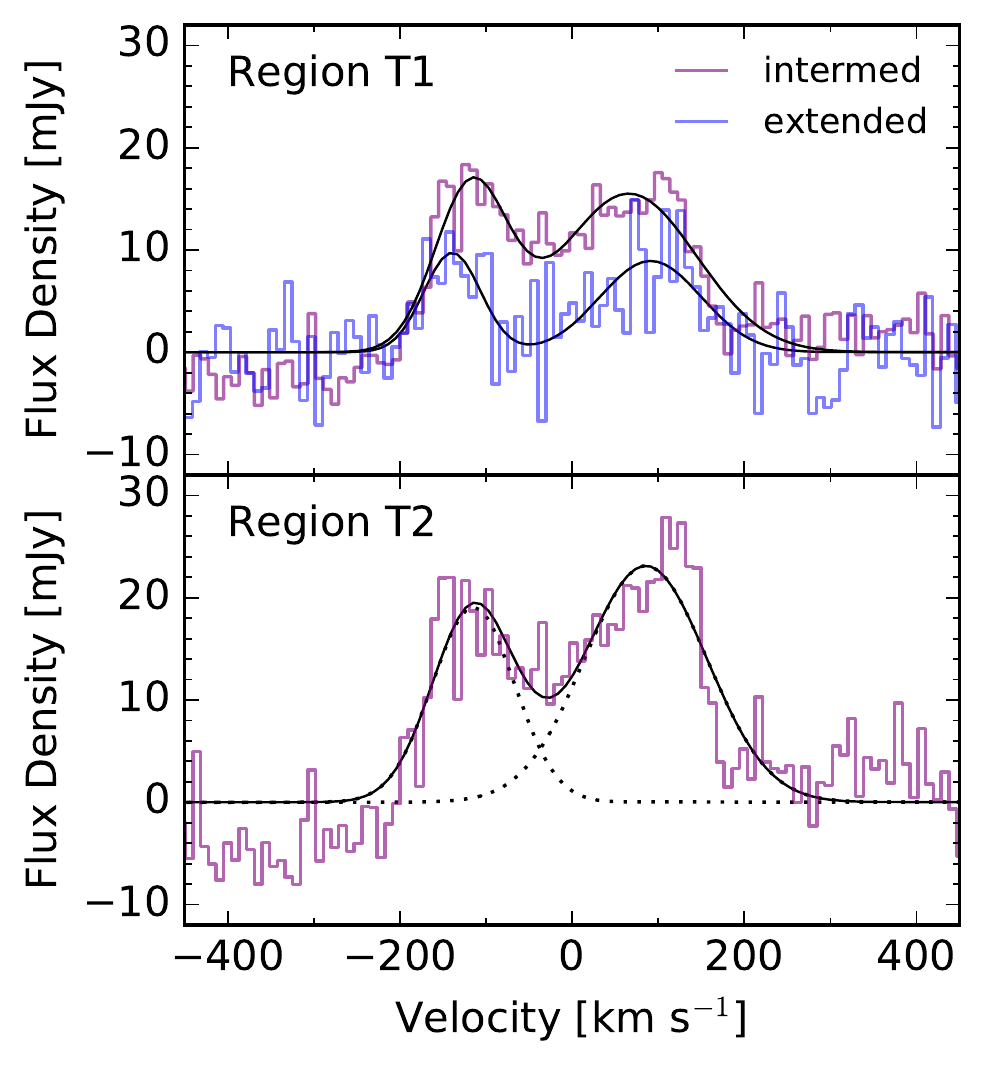}
    \caption{H40$\alpha$ line spectra extracted from the aperture regions T1 (top) and T2 (bottom; see Figure~\ref{fig:totalap}). In blue is the extended-configuration (``extended'') spectrum extracted from the high-resolution 0.2\arcsec\ data; this spectrum shows the maximum total integrated line flux extracted. In purple, the intermediate-configuration (``intermed'') spectra extracted from low-resolution, native 0.7\arcsec\ data; the spectrum from the T2 region is the total maximum integrated line flux from this data. The solid black line represents the sum total of two Gaussian fits. The dotted, black line represents the single Gaussian fits.}
    \label{fig:totalspec}
\end{figure}

In the top panel of Figure~\ref{fig:totalspec}, we show the total recombination line emission extracted from the starburst region in the 0.2\arcsec\ resolution, ``extended'' configuration observations. The aperture we use, designated as region T1, is shown in Figure~\ref{fig:totalap}. Details of the aperture selection are described in Section~\ref{sec:intermedtot}. The spectrum consists of two peaks reminiscent of a double horn profile representing a rotating ring. We fit the spectrum using the sum of two Gaussian components. In Table~\ref{tab:totalline}, we include the properties of the best fit line profiles. The sum total area of the fits is ($2.1\pm0.6$)~Jy~\kms. 

\begin{deluxetable*}{c c C C C C C C C }
\tablecaption{H40$\alpha$ line profiles from the regions of total flux 
\label{tab:totalline}}
\tablewidth{0pt}
\tablehead{
\colhead{Region} %1
& \colhead{Config} %2
& \colhead{v$_{\mathrm{cen},1}$} %3
& \colhead{Peak$_1$} %4
& \colhead{FWHM$_1$} %5
& \colhead{v$_{\mathrm{cen},2}$} %6
& \colhead{Peak$_2$} %7
& \colhead{FWHM$_2$} %8
& \colhead{H40$\alpha$ flux} %9
\\
\colhead{} & \colhead{} & \colhead{(\kms)} & \colhead{(mJy)} & \colhead{(\kms)} & \colhead{(\kms)} & \colhead{(mJy)} & \colhead{(\kms)} & \colhead{(mJy~\kms)}
}
\startdata
T1 &  extend & -139 \pm 9 & 9.7 \pm 2 & 78 \pm 22 & 91 \pm 13 & 8.9 \pm 1.8 & 138 \pm 14 & 2100 \pm 600 \\
T1 &  intermed & -117 \pm 9 & 16 \pm 3 & 99 \pm 21 & 66 \pm 2 & 16 \pm 2 & 186 \pm 15 & 4800 \pm 900 \\ 
T2 &  intermed & -114 \pm 13 & 19 \pm 4 & 113 \pm 31 & 86 \pm 13 & 23 \pm 3 & 167 \pm 14 & 6400 \pm 1000 \\ 
\enddata
\end{deluxetable*}

%%%%%%%%%%%%%%%%%%%%%%%%%%%%%%%%%%%%%%%%

\subsection{Line Emission from 0.7 \arcsec\ resolution, Intermediate-configuration Observations}
\label{sec:intermedemission}

Figure~\ref{fig:totalspec} also shows integrated spectra derived from intermediate-resolution (0.7\arcsec) data. We use these data as a tracer of the total ionizing photons of the starburst region. We expect that the intermediate-resolution data includes emission from both discrete, point-like sources and diffuse emission from any smooth component. 

The total integrated emission in the intermediate-configuration data is about three times larger than the integrated emission in the extended-configuration data. Spectra representing the total integrated line flux are shown in Figure~\ref{fig:totalspec}.
In Figure~\ref{fig:totalap}, we show the integrated intensity map of H40$\alpha$ emission from the intermediate-configuration (0.7\arcsec) data. In Table~\ref{tab:totalline}, we include the best fit line profiles. 

We also compare the line profiles of H40$\alpha$ and H42$\alpha$ in the intermediate-configuration (0.7\arcsec) data, see Table~\ref{tab:intermedline} and Figures~\ref{fig:B16ap}~\&~\ref{fig:B16spec}. We find that the integrated line emission of H42$\alpha$ is enhanced compared with H40$\alpha$, reaching a factor of 2 greater when integrated over the entire starburst region. Yet we see good agreement between the two lines at the scale of individual cluster candidates. Spectral lines (possibly arising from $c$-C$_3$H$_2$) likely contaminate the H42$\alpha$ line flux in broad, typically spatially unresolved, line profiles.

%%%%%%%%%%%%%%%%%%%%%%%%%%%%%%%%%%%%%%%%

\subsubsection{Total Emission from the Starburst Region}
\label{sec:intermedtot}

\begin{figure}
    \centering
    \includegraphics[width=0.47\textwidth]{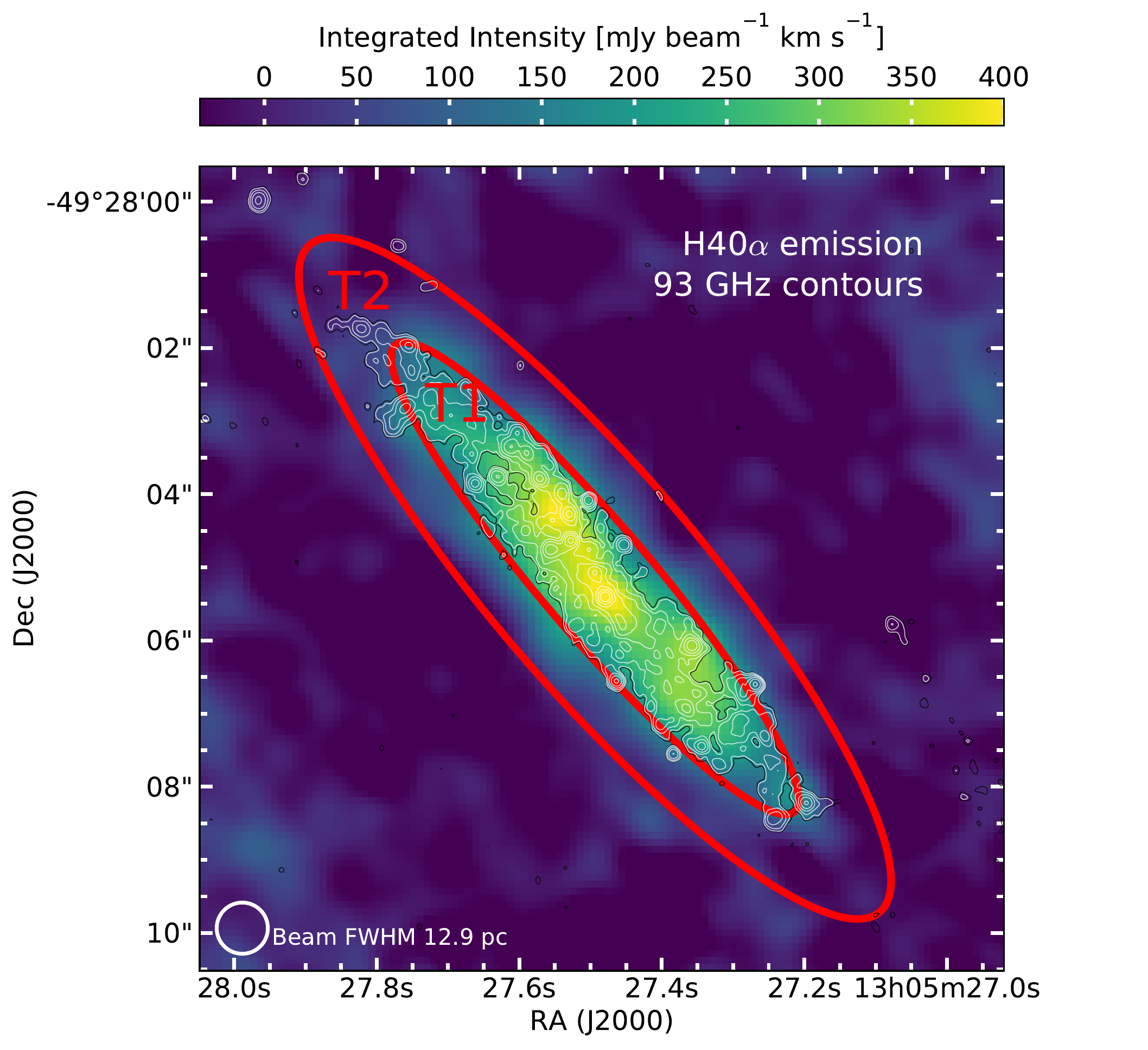}
    \caption{Integrated intensity (moment 0) map of H40$\alpha$ emission integrated between $V_{\mathrm{systemic}} \pm 170$~\kms\ and observed with the intermediate telescope configuration at native 0.7\arcsec\ resolution. Overlaid are contours of the 93 GHz continuum from extended-configuration, high-resolution (0.12\arcsec) data – as described in Figure 1. Red ellipses mark the apertures used to extract the total line emission from regions T1 and T2. }
    \label{fig:totalap}
\end{figure}

% introduce Figure 8
In Figure~\ref{fig:totalap} we show the integrated intensity map of H40$\alpha$ emission integrated between $\mathrm{v_{systemic}} \pm 170$~\kms, as calculated from the 0.7\arcsec\ intermediate-configuration data. Diffuse emission is detected throughout the starburst region and up to 30~pc in apparent size beyond the region where we detect the bright point sources at high resolution.

Also shown in Figure~\ref{fig:totalap} are the apertures used to extract spectra in Figure~\ref{fig:totalspec}. We fit a two dimensional Gaussian to the continuum emission in the 0.7\arcsec\ resolution observations (see Figure~\ref{fig:B16ap}). This results in a best fit centered at $(\alpha, \delta) = (13\mathrm{h}\, 05\mathrm{m}\, 27.4896\mathrm{s}, -49^{\circ}\, 28\arcmin\, 05.159\arcsec)$, with major and minor Gaussian widths of $\sigma_{\mathrm{maj}} = 2.4$\arcsec\ and $\sigma_{\mathrm{min}} = 0.58$\arcsec, and an angle of $\theta = 49.5^{\circ}$; we use this fit as a template  for the aperture location, position angle, width and height.  We independently vary the major and minor axes (in multiples of $0.5\sigma_{\mathrm{maj}}$ and $0.5\sigma_{\mathrm{min}}$, respectively) in order to determine the aperture which maximizes the total integrated signal in channels within $\pm 170$~\kms. With the extended-configuration cube, we find the largest integrated line emission with an aperture of 8.4\arcsec~$\times$~1.5\arcsec, which we refer to as T1. With the intermediate-configuration cube, the largest integrated line emission arises with an aperture of 12.0\arcsec~$\times$~2.9\arcsec, which we refer to as T2.

We extract the total H40$\alpha$ line flux from the intermediate-configuration (0.7\arcsec\ resolution) cube using the T2 aperture (see bottom panel of Figure~\ref{fig:totalspec}). The spectrum shows a double horn profile, indicating ordered disk-like rotation. We fit the features with two Gaussians. The sum total of their integrated line flux is ($6.4 \pm 1.0$)~Jy~\kms.

We also extract H40$\alpha$ line flux from the intermediate-configuration (0.7\arcsec\ resolution) cube within the T1 aperture in order to directly compare the integrated line flux in the two different data sets using the same aperture regions. We find more emission in the intermediate-configuration data, a factor of $\sim$2.3 greater than the extended-configuration data. This indicates that some recombination line emission originates on large scales ($>$100~pc) to which the high-resolution, long baselines are not sensitive.

%%%%%%%%%%%%%%%%%%%%%%%%%%%%%%%%%%%%%%%%

\subsubsection{H42$\alpha$ Contamination}
\label{sec:42a_contam}

In this section we compare the line profiles from H40$\alpha$ and H42$\alpha$ extracted from the 0.7\arcsec\ intermediate configuration data. In principle, we expect the spectra to be virtually identical, which is why we average them to improve the signal-to-noise at high-resolution. Here we test that assumption \edit1{at low-resolution}. To summarize, we find evidence that a spectral line may contaminate the H42$\alpha$ measured line flux in broad (typically spatially unresolved) line profiles. Yet we see good agreement between the two lines at the scale of individual cluster candidates.

\begin{figure}
    \centering
    \includegraphics[width=0.47\textwidth]{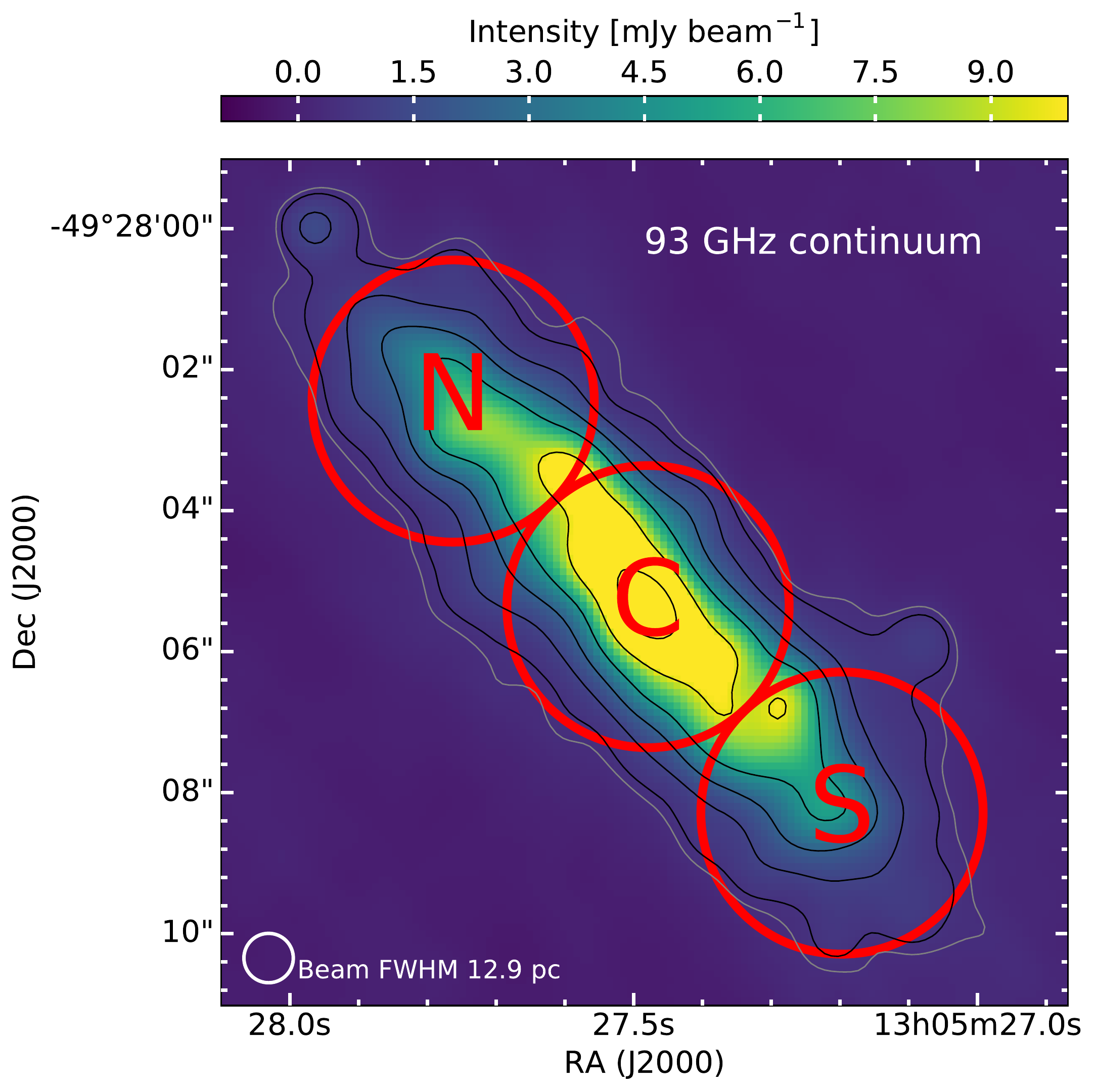}
    \caption{Continuum emission at 93~GHz observed with an intermediate configuration with native resolution $\mathrm{FWHM} = 0.7$\arcsec\ (or 12.9 pc at the distance of \trg).  The rms noise away from the source is $\sigma \approx 0.15$~\mJyb. Contours of the continuum image show $3\sigma$ emission (gray) and $[4,8,16,...]\sigma$ emission (black). Apertures (red) with a diameter of 4\arcsec\ mark the regions N, C, and S. }
    \label{fig:B16ap}
\end{figure}

We extracted spectra in three apertures to demonstrate the constant velocity offset of the contaminants. We approximately matched the locations of these apertures to those defined in \cite{Bendo2016}, in which H42$\alpha$ was analyzed at 2.3\arcsec\ resolution; in this way we are able to confirm the flux and line profiles we extract at 0.7\arcsec\ resolution with those at 2.3\arcsec. The non-overlapping circular apertures with diameters of 4\arcsec\ designated as North (N), Center (C), and South (S) are shown in Figure~\ref{fig:B16ap}. 

\begin{figure}
    \centering
    \includegraphics[width=0.44\textwidth]{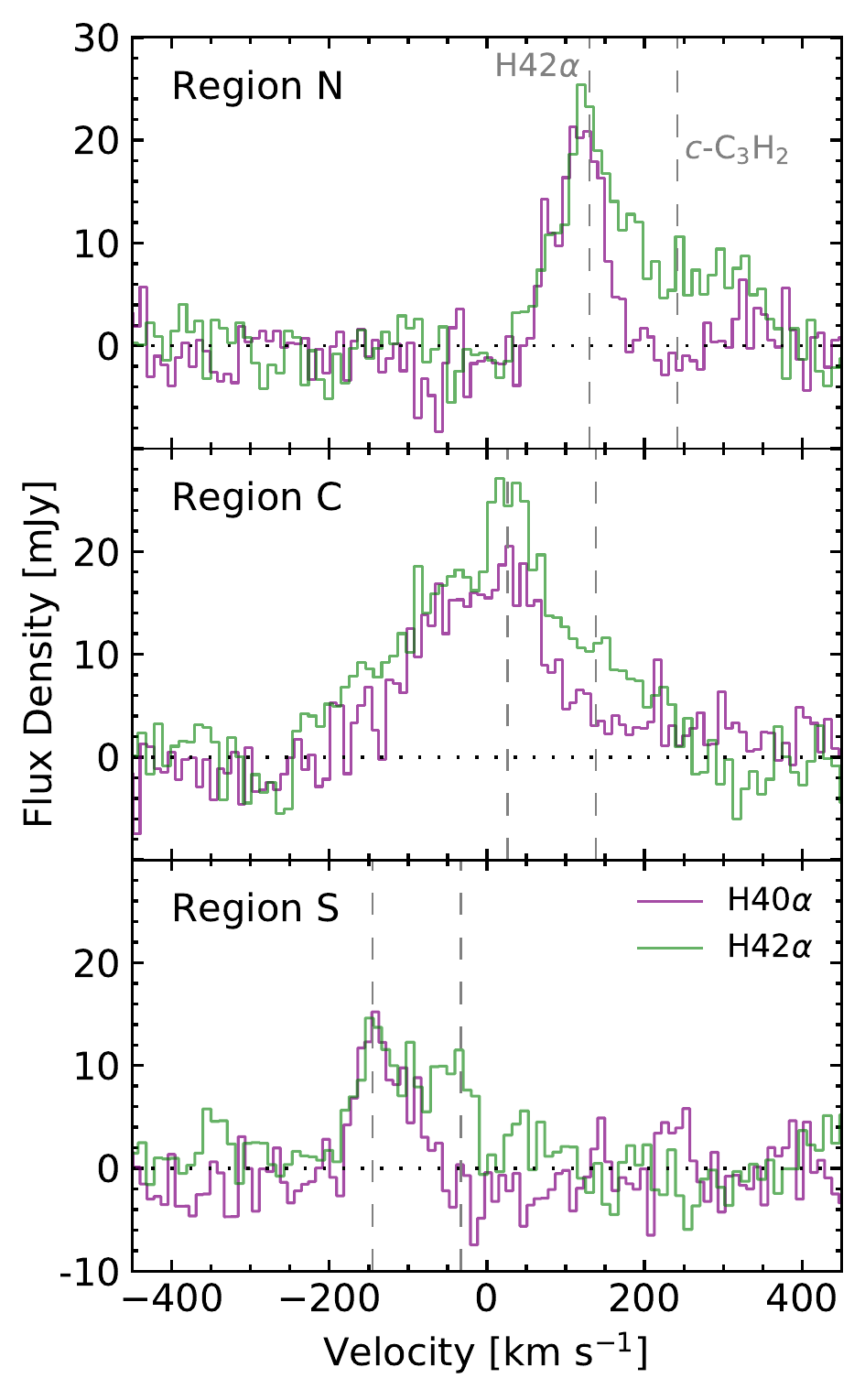}
    \caption{Comparison of our H40$\alpha$ (purple) and H42$\alpha$ (green) from intermediate configuration, low-resolution data from the regions defined in Figure~\ref{fig:B16ap} as N (top), C (middle), and S (bottom). We find H42$\alpha$ to be contaminated by spectral lines which may include $c$-C$_3$H$_2$ $4_{32} - 4_{23}$ --- shown as a dashed line in the panels at expected velocities with respect to H42$\alpha$. When contaminant lines are included, the integrated line flux of H42$\alpha$ is over estimated by a factor of 1.5 in these apertures; this grows to a factor of 2 when integrating over the total starburst emission.}
    \label{fig:B16spec}
\end{figure}

We used our intermediate-configuration data to extract an H40$\alpha$ and an H42$\alpha$ spectrum in each of the regions. We overplot the spectra of each region in Figure~\ref{fig:B16spec}. \edit1{We fit a single Gaussian profile to the line emission, except for H42$\alpha$ emission in region N where two Gaussian components better minimized the fit.} The total area of the fits are presented in Table~\ref{tab:intermedline} as the integrated line flux. 

Our line profiles of H42$\alpha$ are similar in shape and velocity structure as those analyzed in \cite{Bendo2016} and the integrated line emission is also consistent (within 2$\sigma$). This indicates that we are recovering the H42$\alpha$ total line flux and properties with our data.

On the other hand, the H40$\alpha$ flux we extract is about a factor of $\sim$1.6 lower than the H42$\alpha$ fluxes in these apertures (see Figure~\ref{fig:B16spec} and Table~\ref{tab:intermedline}). The discrepancy grows to a factor of 2 in the profile extracted from the total region. 

\begin{deluxetable}{c C C C }
    \tablecaption{Comparison of integrated recombination line flux \label{tab:intermedline}}
    \tablewidth{0pt}
    \tablehead{
    \colhead{Region} %1
    & \colhead{H42$\alpha$ flux} %2
    & \colhead{H40$\alpha$ flux} %3
    & \colhead{Ratio H42$\alpha$/H40$\alpha$} \\ %4
    \colhead{}  & \colhead{(Jy~\kms)} & \colhead{(Jy~\kms)} &
    \colhead{} }
    \startdata
N & 2.1 \pm 0.2 & 1.5 \pm 0.2  & 1.4 \pm 0.2 \\ 
C & 5.9 \pm 0.3 & 3.7 \pm 0.3 & 1.6 \pm 0.2 \\
S & 1.9 \pm 0.2 & 0.96 \pm 0.1 & 2.0 \pm 0.3 \\ 
    \enddata
\end{deluxetable}

The additional emission seen in the H42$\alpha$ spectrum at velocities +100 \kms\ to +250 \kms\ with respect to the bright, presumably hydrogen recombination line peak, is absent in the H40$\alpha$ profile. It is not likely to be a maser-like component of hydrogen recombination emission since the relative flux does not greatly vary in different extraction regions, and densities outside of the circumnuclear disk would not approach the emission measures necessary (e.g., $EM_v \gtrsim 10^{10}$~cm$^{-6}$~pc$^3$) for stimulated line emission. 

We searched for spectral lines in the frequency range $\nu_{\mathrm{rest}} \sim 85.617$ -- 85.660 GHz, corresponding to these velocities and find several plausible candidates, though we were not able to confirm any species with additional transitions in the frequency coverage of these observations. A likely candidate may be the $4_{32} - 4_{23}$ transition of $c$-C$_3$H$_2$. $c$-C$_3$H$_2$ has a widespread presence in the diffuse ISM of the Galaxy \citep[e.g.,][]{Lucas2000} and the $2_{20} - 2_{11}$ transition has been detected in \trg\ \citep{Eisner2019}.
As an example we plot the velocity of $c$-C$_3$H$_2$ $4_{32} - 4_{23}$ relative to H42$\alpha$ in Figure~\ref{fig:B16spec}.

%%%%%%%%%%%%%%%%%%%%%%%%%%%%%%%%%%%%%%%%

\section{Physical Properties of the Candidate Star Clusters}    
\label{sec:properties}

In this section, we estimate properties of the candidate star clusters, summarized in Table~\ref{tab:props}. We discuss their size and approximate age. Properties of the ionized gas content, such as temperature (see Table~\ref{tab:temp}), metallicity, density and mass are derived from the continuum and recombination line emission. We estimate the ionizing photon rate of the candidate stars clusters and use it to infer the stellar mass (see Figure~\ref{fig:cmf}). From the dust emission at 350~GHz, we estimate gas masses of the candidate star clusters. With a combined total mass from gas and stars, we estimate current mass surface densities and free-fall times. 

We exclude Source 5 from the analysis since the free-free fraction is $f_{\mathrm{ff}} < 0.01$. We also remove the presumed AGN core (Source 18) from the analysis.

%%%%%%%%%%%%%%%%%%%%%%%%%%%%%%%%%%%%%%%%

\subsection{Size}    
\label{sec:size}

The sources identified through \texttt{PyBDSF} in the 93~GHz continuum image are fit with two dimensional Gaussians. The average of the major and minor (convolved) FWHM is listed in Table~\ref{tab:props} as the FWHM size of the source in units of pc. The Gaussian fits are all consistent with circular profiles within error. FWHM sizes of (1.4--4.0)~pc are observed, consistent with typical sizes of young, massive star clusters \citep{Ryon2017,Leroy2018}. However, the lower end may reflect the resolution limit of our beam, with a FWHM size of 2.2~pc. The uncertainties we report reflect the errors of the Gaussian fit. 

Based on high-resolution imaging of embedded clusters in the nucleus of the Milky Way and NGC~253, some of these clusters might break apart at higher resolution \citep[][Levy et al., in prep]{Ginsburg2018a}. If they follow the same pattern seen in these other galaxies, each source would have one or two main components potentially with several associated fainter components.

%%%%%%%%%%%%%%%%%%%%%%%%%%%%%%%%%%%%%%%%

\subsection{Age} 
\label{sec:age}

Throughout our analysis, we assume that the candidate star clusters formed in an instantaneous burst of star-formation roughly 5~Myr ago (with a likely uncertainty of $\sim$1~Myr). This (approximately uniform) age is supported through the coincident detection of RRLs and supernovae remnants, previous analyses of the global population of the burst \cite[e.g.,][]{Marconi2000, Spoon2000} and an orbital timescale of $\approx$3~Myr for the starburst region. We elaborate on this supporting evidence below.

As we discuss in Section~\ref{sec:freefreefrac}, dust does not significantly contribute to 93~GHz emission \edit1{(with a median fraction of $f_{\mathrm{d}} = 0$)}, but synchrotron emission does through supernova remnants. Supernova explosions begin from $\sim$3~Myr in the lifetime of a cluster and cease around $\sim$40~Myr when the most massive stars have died out; this puts the loosest bounds on the age of the candidate clusters we observe. The coincident detection of supernovae remnants in a third (6/15) of our recombination line detected sources implies that the burst is likely not at the earliest stage of the supernovae phase. However the ionizing photon rate changes dramatically over 3~Myr to 10~Myr, dropping by about two orders of magnitude \citep{Leitherer1999}. As a result clusters are significantly harder to detect in radio recombination lines or free-free continuum emission after $\sim$5~Myr. 

Properties of star-forming activity in the central starburst have been estimated by combining far-infrared (FIR) and optical/IR tracers. \cite{Marconi2000} discerned an age of 6 Myr and mass of $4\times 10^{7}$~\Msun\ by using Pa$\alpha$ and Br$\gamma$ to trace the energy distribution of the photon output of the population. However, the dust extinction was underestimated, complicated by the uncertainty in the AGN contribution.  Mid-infrared (MIR) observations with the Infrared Space Observatory \citep[ISO; ][]{Kessler1996} of line ratios further constrained this scenario. \cite{Spoon2000} estimated an extinction of $A_{\mathrm{V}} = 36^{+18}_{-11}$, determined that the AGN is not dominating the ionizing radiation field, and found that the star-forming population is consistent with a burst of age $\geq$5~Myr. 

As a sanity check on whether a synchronized burst might be expected, we calculate the orbital timescale associated with the the burst region. Taking the rotation velocity $\sim 170$~km~s$^{-1}$ from the integrated spectrum and the radius $\sim 80$~pc associated with region T1, we estimate an orbital timescale of $\sim 3$~Myr. If we take this as roughly the timescale for the nuclear disk to react to changing conditions, a burst shutting off or turning on in a $\sim 5$~Myr timescale is reasonable.

%%%%%%%%%%%%%%%%%%%%%%%%%%%%%%%%%%%%%%%%

\subsection{Temperature and Metallicity}    
\label{sec:temp}

The ratio of the integrated recombination line flux (Equation~\ref{eq:Sline}) to the free-free continuum flux density (Equation~\ref{eq:Scont}) allows the electron temperature to be determined. Dependencies on the distance, emission measure, and (possible) beam-filling effects cancel out under the assumption that the two tracers arise in the same volume of gas. We show in Appendix~\ref{ap:temp}, that when taking the ratio of the integrated line to continuum, $R_{\mathrm{LC}}$, and solving for the temperature, $T_e$, we arrive at
\begin{equation}
\begin{split}
T_e = 10^4~\mathrm{K}
	\left[ b_{\mathsf{n+1}}	\left( 1 + y \right)^{-1} 
	\left( \frac{ R_{\mathrm{LC}} }{ 31.31~\mathrm{km~s^{-1}} } \right)^{-1} \right. \times \\
    \left. \left( \frac{\nu}{100~\mathrm{GHz}} \right)^{1.12} \right]^{0.85}
\label{eq:temp}
\end{split}
\end{equation}
where $b_{\mathsf{n+1}}$ is the non-LTE departure coefficient, and $y$ is the abundance ratio of ionized helium to hydrogen number density, $y= n_{He+} / n_p$, which we fix as $y = 0.10$ \citep{DePree1996, Mills2020}. 

Table~\ref{tab:temp} lists the temperatures we derive in the region. We focus on the 5 sources with bright (peak S/N $>4.7\sigma$) and well-fit recombination line emission. \edit1{Most of these sources have higher free-free fractions than the median.} To derive the temperatures, we re-evaluate the continuum (fraction of) free-free emission at the resolution of 0.2\arcsec, since the free-free fraction may change with resolution.  Therefore, we convolve the Band 3 continuum images to 0.2\arcsec\ resolution. We extract the continuum from the full-bandwidth image through aperture photometry, using an aperture diameter of 0.4\arcsec. In order to exactly match the processing of the spectral line data, we do not subtract background \edit1{continuum} emission within an outer annulus. Then, by extracting the continuum in each spectral window (using the same aperture diameters just described), we fit for the in-band spectral index. We use the procedure described in Section~\ref{sec:freefreefrac} to constrain the free-free fraction from the spectral index fit. 

\begin{deluxetable}{c C C C C C}
\tablecaption{Temperature analysis \label{tab:temp} }
\tablewidth{0pt}
\tablehead{
\colhead{Source} %1
& \colhead{$\int S_{\mathrm{L}}\,\mathrm{dV}$} %2
& \colhead{$S_{93}$} %3
& \colhead{$f_{\mathrm{ff}}$} %4
& \colhead{$T_e$} \\ %6
\colhead{} & \colhead{(mJy~\kms)} & \colhead{(mJy)} & \colhead{} & \colhead{(K)}
}
\startdata
08 &  92 \pm  15 & 3.5 \pm 0.4 & 0.74 \pm 0.20 & 6000 \pm 1700  \\ 
13 &  99 \pm  14 & 2.9 \pm 0.3 & 0.91 \pm 0.22 & 5600 \pm 1400  \\ 
22 & 100 \pm  21 & 3.4 \pm 0.3 & 0.78 \pm 0.20 & 5600 \pm 1700  \\ 
26 &  61 \pm  13 & 2.6 \pm 0.3 & 0.72 \pm 0.29 & 6400 \pm 2600  \\ 
27 &  50 \pm  10 & 3.6 \pm 0.4 & 0.45 \pm 0.16 & 6500 \pm 2300  \\ 
\enddata
\tablecomments{$\int S_{\mathrm{L}}\,\mathrm{dV}$ refers to the integrated line emission. $S_{93}$ is the continuum flux density extracted at 93~GHz in the 0.2\arcsec\ resolution image. $f_{\mathrm{ff}}$ is the estimated free-free fraction at 0.2\arcsec\ resolution. $T_e$ is the electron temperature derived using Equation~\ref{eq:temp}. }
\end{deluxetable}

With the free-free fraction and measured fluxes, we plug in the line to continuum ratio into Equation~\ref{eq:temp} and take $b_{\mathsf{n}} = 0.73$ \citep{Storey1995} to arrive at the temperature. The departure coefficient at $\mathsf{n} = 41$ is loosely ($<15$\% variation) dependent on the temperature. We iterate (once) on the input $b_{\mathsf{n}}$ and output temperature. $b_{\mathsf{n}} = 0.73$ is the modeled value for this temperature and for typical densities of $n_e = (10^3 - 10^4)$~cm$^{-3}$ of ionized gas surrounding young, massive stars and consistent with the ionized gas densities we derive in Section~\ref{sec:iongas}.

The uncertainties in the electron temperatures we derive in Table~\ref{tab:temp} are dominated by the uncertainties in the free-free fraction. We take the mean and standard deviation values of $T_e = (6000 \pm 400)$~K as a representative electron temperature of the ionized plasma in the candidate star clusters. This temperature is consistent with the temperature derived from a lower-resolution analysis of \trg\ at 2.3\arcsec~$\times$~2.6\arcsec\ resolution, which finds $T_e = (5400 \pm 600)$~K \citep{Bendo2016}.

Our estimated temperature implies a thermal line width of $(16 \pm 4)$~\kms\ \citep{Brocklehurst1972}. Given that this is smaller than our observed line widths, non-thermal motions from bulk velocities (such as turbulence, inflow or outflow) must contribute to broadening the spectral line profiles.

The electron temperature of free-free plasma surrounding massive stars is related to the metallicity of the plasma, as the metals contribute to gas cooling. \cite{Shaver1983} established a relation,  
\begin{equation}
\begin{split}
12 + \log_{10}(\mathrm{O/H}) = (9.82 \pm 0.02) - \\ (1.49\pm 0.11) \frac{T_e}{10^4\,\mathrm{K}},
\end{split}
\end{equation}
with the temperatures and metallicities derived with (auroral) collisionally excited lines at optical wavelengths. Furthermore, they showed that these temperatures are consistent with electron temperatures derived from radio recombination lines. We find a representative O/H metallicity of $12 + \log_{10}(\mathrm{O/H}) = 8.9 \pm 0.1$. This value is in approximate agreement (within 2$\sigma$) with the average metallicity and standard deviation of $12 + \log_{10}(\mathrm{O/H}) = 8.5 \pm 0.1$ \citep{Stanghellini2015} determined in 15 star-forming regions in the galactic plane of \trg\ (and which is consistent with no radial gradient) using strong-line abundance ratios of oxygen, sulfur, and nitrogen spectral lines.

%%%%%%%%%%%%%%%%%%%%%%%%%%%%%%%%%%%%%%%%

\subsection{Ionized Gas: Emission Measure, Density and Mass} \label{sec:iongas}

We determine the volumetric emission measure of gas ionized in candidate stars clusters using Equations~\ref{eq:Scont} and \ref{eq:Sline} together with the mean temperature derived in Section~\ref{sec:temp}. In Table~\ref{tab:props}, we list the results for each candidate star cluster. Emission measures that we determine from the free-free continuum range from 
$\log_{10} (EM_{\mathrm{C}} / \mathrm{cm^{-6} \, pc^3}) \sim 7.3$ -- 8.7, with a median value of 8.4. We also calculate the volumetric emission measure of ionized hydrogen as determined by the effective H41$\alpha$ recombination line when applicable, noting that $EM_{\mathrm{C}} = (1+y)\,EM_{\mathrm{L}}$. The line emission measures
range from 
$\log_{10} (EM_{\mathrm{L}} / \mathrm{cm^{-6} \, pc^3}) \sim 8.4$ -- 8.9, with a median value of 8.5. The uncertainty in the emission measures is $\sim$0.4 dex and is dominated by the errors of the free-free fraction.

Next, we solve for the electron density. We use the emission measure determined from the free-free continuum, assume $n_e = n_+$, and consider a spherical volume with $r = \mathrm{FWHM_{size}} / 2$. We arrive at densities between $\log_{10} (n_e/\mathrm{cm}^{-3})= 3.1$--3.9 with a median value of 3.5. 

We matched (see Section~\ref{sec:fluxextract}) five of the candidate star clusters that have recombination line emission detected -- Sources 1, 6, 14, 21, 27 -- with the 2.3~GHz objects of \cite{Lenc2009} which have the free-free optical depth modeled through their low-frequency turnovers. Although the 2.3~GHz objects have non-thermal indices, it is their radio emission which is opaque to free-free plasma. Using the optical depths derived in \cite{Lenc2009} and our fiducial electron temperature, we solve for the density through the relation, $\tau \approx 3.28 \times 10^{-7} \left( \frac{T_e}{10^4~\mathrm{K}} \right)^{-1.35} \left( \frac{\nu}{\mathrm{GHz}} \right)^{-2.1} \left( \frac{EM_{\ell}}{\mathrm{cm^{-6}~pc}} \right)$ \citep{Condon2016}, where $EM_{\ell} = n_e n_+ \ell$ and for which a spherical region the pathlength $\ell$ translates as $\ell = \frac{3}{4}r$. We find densities in the range $\log_{10} (n_e/\mathrm{cm}^{-3})= 3.3$~--~3.6. This agrees well with the values we separately derive.

We convert the ionized gas density and source sizes to an ionized gas mass through,
\begin{equation}
M_{+} = 1.36 m_{\mathrm{H}} \, n_+  \frac{4}{3}\pi r^3
\end{equation}
where we have assumed a 1.36 contribution of helium by mass and we let $r = \mathrm{FWHM_{size}} / 2$. The ionized gas masses of the candidate star clusters range from $\log_{10} (M_+$ / \Msun ) = 2.7 -- 3.5 with a median value of 3.1. The ionized gas mass is a small fraction ($\lesssim 1$\%) of the stellar mass (see Section~\ref{sec:mass}).

%%%%%%%%%%%%%%%%%%%%%%%%%%%%%%%%%%%%%%%%

\subsection{Ionizing Photon Production and Stellar Mass}    
\label{sec:mass}

\begin{deluxetable*}{c C C C C C C C C}
\setlength{\tabcolsep}{8pt}
\tablecaption{Physical properties of candidate star clusters \label{tab:props}}
\tablewidth{0pt}
\tablehead{
\colhead{Source} %1
& \colhead{FWHM} %2
& \colhead{$ \log{ (EM_{\mathrm{C}}) }$ $^{a}$ } %3
& \colhead{$ \log{ (EM_{\mathrm{L}}) }$ $^{a}$ } %3
& \colhead{$ \log{ (Q_0) } $ $^{a}$} %4
& \colhead{$ \log{ (M_{\star}) } $ $^{a}$} %5
& \colhead{$ \log{ (M_{\mathrm{gas}}) } $ $^{a}$} %5
& \colhead{$ \log{ (\Sigma_{\mathrm{Tot}}) } $ $^{a}$} %5
& \colhead{$ \log{ (t_{\mathrm{ff}}) } $ $^{a}$} \\ %5
\colhead{} & \colhead{(pc)} & \colhead{(cm$^{-6}$~pc$^3$)} & \colhead{(cm$^{-6}$~pc$^3$)} & \colhead{(s$^{-1}$)} & \colhead{(\Msun)}
& \colhead{(\Msun)} & \colhead{(\Msun~pc$^{-2}$)}  & \colhead{(yr)}}
\startdata
01 & 3.0 \pm 0.1 & 8.1 & 8.6 & 51.2 & 5.1 & <4.5 & 4.0 & 4.9  \\ 
02 & 2.6 \pm 0.1 & 8.2 & ... & 51.3 & 5.2 & 4.8 & 4.3 & 4.7  \\ 
03 & 3.1 \pm 0.1 & 8.2 & 8.4 & 51.2 & 5.2 & 4.8 & 4.1 & 4.9  \\ 
04 & 2.5 \pm 0.1 & 7.7 & 8.5 & 50.8 & 4.7 & 4.6 & 4.0 & 4.9  \\ 
06 & 2.9 \pm 0.1 & 8.0 & 8.4 & 51.1 & 5.0 & <4.5 & 3.8 & 5.0  \\ 
07 & 2.7 \pm 0.1 & 8.1 & 8.4 & 51.2 & 5.1 & 4.6 & 4.2 & 4.8  \\ 
08 & 2.4 \pm 0.1 & 8.5 & 8.7 & 51.5 & 5.5 & 4.5 & 4.5 & 4.6  \\ 
09 & 2.4 \pm 0.1 & 8.1 & ... & 51.2 & 5.1 & <4.3 & 4.1 & 4.8  \\ 
10 & 2.9 \pm 0.1 & 8.2 & ... & 51.2 & 5.1 & <4.5 & 4.0 & 4.9  \\ 
11 & 1.4 \pm 0.1 & 8.0 & ... & 51.1 & 5.0 & 4.0 & 4.6 & 4.5  \\ 
12 & 2.7 \pm 0.1 & 8.2 & 8.7 & 51.3 & 5.2 & 5.0 & 4.4 & 4.7  \\ 
13 & 2.5 \pm 0.1 & 8.5 & 8.7 & 51.6 & 5.5 & 4.7 & 4.6 & 4.6  \\ 
14 & 2.2 \pm 0.1 & 8.1 & 8.9 & 51.2 & 5.1 & 4.4 & 4.3 & 4.7  \\ 
15 & 2.5 \pm 0.1 & 8.0 & 8.4 & 51.0 & 5.0 & <4.4 & 4.0 & 4.9  \\ 
16 & 2.3 \pm 0.1 & 8.1 & ... & 51.2 & 5.1 & 4.7 & 4.3 & 4.7  \\ 
17 & 3.1 \pm 0.1 & 8.7 & ... & 51.7 & 5.7 & 5.2 & 4.6 & 4.6  \\ 
19 & 2.4 \pm 0.1 & 7.4 & ... & 50.5 & 4.4 & <4.3 & 3.5 & 5.1  \\ 
20 & 3.7 \pm 0.1 & 8.4 & ... & 51.4 & 5.3 & 5.1 & 4.2 & 4.9  \\ 
21 & 3.1 \pm 0.1 & 7.3 & 8.8 & 50.4 & 4.3 & <4.6 & 3.1 & 5.4  \\ 
22 & 2.4 \pm 0.1 & 8.7 & 8.7 & 51.8 & 5.7 & 4.9 & 4.8 & 4.5  \\ 
23 & 2.6 \pm 0.1 & 8.2 & ... & 51.3 & 5.2 & 4.4 & 4.3 & 4.7  \\ 
24 & 3.1 \pm 0.1 & 7.6 & ... & 50.6 & 4.6 & <4.6 & 3.4 & 5.2  \\ 
25 & 3.9 \pm 0.2 & 8.2 & 8.5 & 51.3 & 5.2 & 5.0 & 4.0 & 5.0  \\ 
26 & 3.4 \pm 0.1 & 8.3 & 8.5 & 51.3 & 5.3 & 4.9 & 4.2 & 4.9  \\ 
27 & 2.2 \pm 0.1 & 8.2 & 8.4 & 51.3 & 5.2 & <4.3 & 4.3 & 4.7  \\ 
28 & 4.0 \pm 0.1 & 8.0 & ... & 51.1 & 5.0 & <4.8 & 3.6 & 5.1  \\ 
29 & 2.9 \pm 0.1 & 8.4 & ... & 51.5 & 5.4 & 4.6 & 4.3 & 4.7  \\ 
\enddata
\tablecomments{Source 5 is not included since its free-free fraction is $f_{\mathrm{ff}} < 0.01$; Source 18 is not included since it is the AGN core. FWHM is the source size as best fit from a Gaussian (to flux that has not been deconvolved); the errors reflect the fit of the Gaussian. $EM_{\mathrm{C}}$ is the free-free emission measure derived from the continuum as in Equation~\ref{eq:Scont}. $EM_{\mathrm{L}}$ is the hydrogen free-free emission measure derived from effective H41$\alpha$ as in Equation~\ref{eq:Sline}; we note $EM_{\mathrm{C}} = (1+y)\,EM_{\mathrm{L}}$. $Q_0$ is the ionizing photon rate derived from $EM_{\mathrm{C}}$ as in Equation~\ref{eq:Q}. \Msun\ is the stellar mass derived from $Q_0$ as in Equation~\ref{eq:M}.\\
$^{a}$ The error on these quantities is $\sim$0.4~dex.
}
\end{deluxetable*}

We estimate the number of the ionizing photons needed per second to maintain the total free-free emitting content (see Table~\ref{tab:props}). From the emission measure of ionized gas and the temperature-dependent recombination coefficient for case B recombination, the rate of ionizing photons (see Appendix~\ref{ap:q}) with $E>13.6$~eV is
\begin{equation}
\begin{split}
Q_0 = \left( 3.8 \times 10^{51}~\mathrm{s}^{-1} \right) \left( \frac{ n_e n_+ V }{ 5 \times 10^8~\mathrm{cm^{-6}~pc^3} } \right) \times \\ \left( \frac{ T_e }{ 10^4~\mathrm{K} } \right)^{-0.83} .
\label{eq:Q}
\end{split}
\end{equation}
where $EM_{\mathrm{C}} = n_e n_+ V$ is the volumetric emission measure of the total ionized gas which we take from the continuum derived emission measure, and $T_e$ is the electron temperature of the ionized gas. Our candidate star clusters have ionizing photon rates in the range $\log_{10} (Q_0 / \mathrm{s}^{-1}) \sim 50.4$ -- 51.8. The sum of the ionizing photon rate over all candidate, massive star clusters is $5.3 \times 10^{52}$~s$^{-1}$. In the top panel of Figure~\ref{fig:cmf}, the ionizing photon rates of the candidate clusters are plotted as complementary cumulative fractions.

We use Starburst99 calculations \citep{Leitherer1999} to infer the stellar mass from the ionizing photon output of a 5 Myr old stellar population, via 
\begin{equation}
    M_{\star} \approx \frac{ Q_0 }{ 4.7 \times 10^{45} } \mathrm{M_{\odot}}.
\label{eq:M}
\end{equation}
We arrive at this value by simulating a single $10^6$~\Msun\ stellar population, with the initial mass function (IMF) of \cite{Kroupa2001}, a maximum stellar mass of 100~M$_{\odot}$, and the default stellar evolution tracks and tuning parameters. Then we divide the ionizing photon output at 5 Myr by the initial mass of the stellar population. We note that this is a rough approximation which has not accounted for the amount of ionizing photons absorbed by dust, mass ejected from the system, and/or enhanced emission from stellar binaries. 

Our candidate star clusters have stellar masses in the range $\log_{10} ( M_{\star} / \mathrm{M_{\odot}} ) \sim$~4.7--6.1 (see Table~\ref{tab:props}) with a median of 5.5. The error on the mass estimate is $\sim$0.4~dex. The sum of the stellar masses of the candidate stars clusters is $\approx 1.1 \times 10^7$~\Msun. In the bottom panel of Figure~\ref{fig:cmf}, the estimated stellar masses of the candidate clusters are plotted as cumulative fractions.

\begin{figure}
    \centering
    \includegraphics[width=0.47\textwidth]{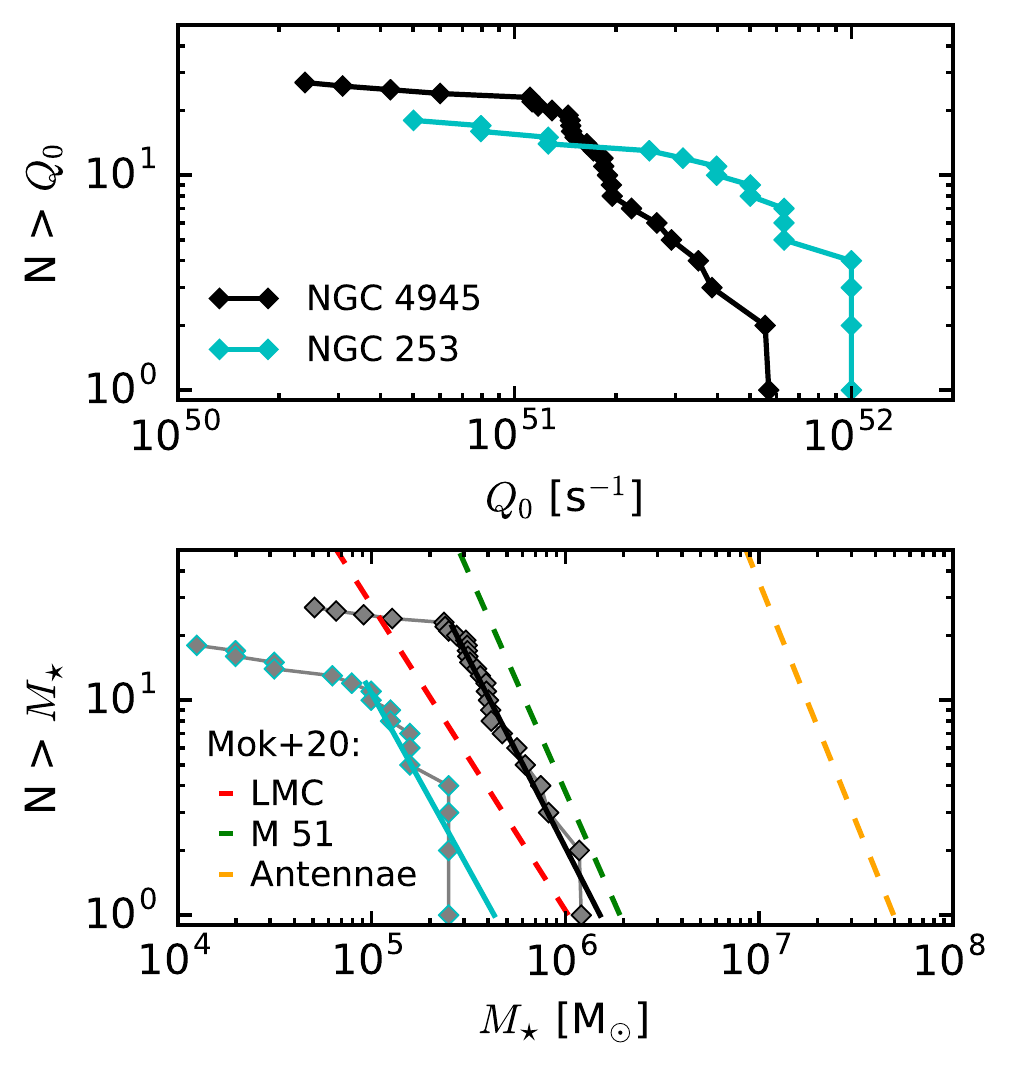}
    \caption{\textit{Top:} The complementary cumulative distribution of the ionizing photon rate, $Q_0$, of candidate star clusters in \trg\ (see Table~\ref{tab:props}) and in NGC~253 \citep{Mills2020}. The turn-off at lower values likely reflects completeness limits. \textit{Bottom:} The stellar mass, $M_{\star}$, of our candidate star clusters inferred from the ionizing photon rates of a cluster with an age of 5~Myr, plotted as a complementary cumulative distribution. We also include the stellar masses of clusters in the starburst of NGC~253 \citep{Mills2020} and in the galaxies LMC, M~51, and the Antennae \citep{Mok2020}. Since the clusters in NGC~253 are likely close to a zero age main sequence, they would produce more ionizing photons per unit mass as compared with the slightly older stellar population in the clusters of NGC~4945.}
    \label{fig:cmf}
\end{figure}

%%%%%%%%%%%%%%%%%%%%%%%%%%%%%%%%%%%%%%%%

\subsection{Gas Mass from Dust} 
\label{sec:dustgas}

We estimate the mass of gas associated with each candidate star cluster (see Table~\ref{tab:props}) from dust emission at 350~GHz.  We determine the dust optical depth by comparing the measured intensity with that expected from an estimate of the true dust temperature. Assuming a mass absorption coefficient, we convert the optical depth to a dust column density. We arrive at a gas mass by multiplying the dust column density with the measured source size and an assumed dust-to-gas mass ratio.

We assume a dust temperature of $T_{\mathrm{dust}} = 130$ K, as has been determined for the gas kinetic temperature in the forming super star clusters in NGC~253 \citep{Gorski2017}. This is an approximation, though the uncertainty is linear. Then we convert the 350 GHz flux density into an intensity ($I_{350}$), and solve for the optical depth through $ I_{350} \approx \tau_{350} B_{\nu}(T_{\mathrm{dust}})$ where $B_{\nu}(T_{\mathrm{dust}})$ is the Planck function evaluated at 350~GHz. We measure optical depths in the range $\tau_{350} \sim $ 0.02 -- 0.10, with a median value of $\tau_{350} \sim 0.04$, justifying our optically thin assumption. We note that the $3\sigma$ upper limit of the sources which have not been detected at 350~GHz corresponds to $\tau_{350}< 0.02$.

Next, we convert the optical depth to a dust column density using an assumed mass absorption coefficient ($\kappa$). We adopt $\kappa = 1.9$~cm$^2$~g$^{-1}$ which should be appropriate for $\nu \sim 350$~GHz and dust mixed with gas at a density of $\sim 10^5 - 10^6$~cm$^{-3}$ \citep{Ossenkopf1994}, but we do note the large (factor of 2) uncertainties on this value. Finally, we combine the dust surface density with an adopted dust-to-gas mass ratio (DGR) of 1-to-100, approximately the Milky Way value and similar to the value found for starburst galaxies by \cite{Wilson2008}. Our estimate for the gas surface density is determined with:
\begin{equation}
\Sigma_{\mathrm{gas}} = \frac{ \tau_{350} }{ \kappa \, \mathrm{DGR} } .
\end{equation}
We determine the gas mass by multiplying the gas surface density by the two dimensional area of the source size, $M_{\mathrm{gas}} = A\, \Sigma_{\mathrm{gas}}$.

The gas masses we estimate are included in Table~\ref{tab:props}. We find values in the range of $\log_{10} (M_{\mathrm{gas}}$ / \Msun ) = 4.4 -- 5.1 with a median value of 4.7. Upper limits for the sources which have not been detected in 350~GHz emission are included in the Table.

%%%%%%%%%%%%%%%%%%%%%%%%%%%%%%%%%%%%%%%%

\subsection{Total Mass from Gas and Stars} 
\label{sec:totalmass}

We estimate the current total mass of each candidate star cluster as the sum of the gas and star masses, where $M_{\mathrm{Tot}} = M_{\mathrm{gas}} + M_{\star}$. The total mass is dominated by the stellar mass, as we find low gas mass fractions of $M_{\mathrm{gas}} / M_{\mathrm{Tot}} = $~0.04~--~0.22 and a median of 0.13. When calculating the total mass of the 10 sources which are not detected at 350~GHz, we do not include the lower limit of the dust mass; we only consider the stellar mass. 

We express the total mass of each source in terms of a surface density (in Table~\ref{tab:props}). This is calculated within the FWHM of the region; we thereby divide the total mass by 2 and divide by the 2D area of the FWHM. The values range from $ \log{ (\Sigma_{\mathrm{Tot}}) } = 3.1$ -- 4.8~\Msun~pc$^{-2}$, or alternatively, $\Sigma_{\mathrm{Tot}} = 0.3$ --13 g~cm$^{-2}$.

Using the total mass, we estimate the gravitational free fall time of the clusters, $t_{\mathrm{ff}} = \left( \frac{ \pi r^3 }{ 8 G M_{\mathrm{Tot}} } \right)^{1/2}$. The values we derive are included in Table~\ref{tab:props}, and range from $\log_{10} (t_{\mathrm{ff}}$ / yr ) = 4.3 -- 5.2  with a median of 4.6. This is the gravitational free fall time that would be experienced by gas with no support if all of the cluster mass were gas. The fact that the gas mass fractions are low and that the age appears longer than the free fall time adds support to the idea that these clusters have mostly already formed.

Summing all sources, we find a total mass within the candidate clusters of $M_{\mathrm{Tot}} \approx 1.5 \times 10^7$~\Msun.

%%%%%%%%%%%%%%%%%%%%%%%%%%%%%%%%%%%%%%%%

\section{Discussion} 
\label{sec:discuss}

Having estimated properties of the candidate star clusters, we explore implications of the results and the role of the candidate clusters with respect to the starburst. 
Before elaborating on that, we discuss the two main sources of uncertainty in the properties of the candidate star clusters: 
the free-free fractions at 93~GHz and the age of the burst.

%%%%%%%%%%%%%%%%%%%%%%%%%%%%%%%%%%%%%%%%

\subsection{Discussion of uncertainties}

Both the free-free fractions at 93~GHz and the age of the burst could be better constrained using future observations, and this would result in more accurate estimates of almost all properties of the cluster candidates. 

The first source of uncertainty that we discuss is the estimate of the free-free flux, by which we estimate a free-free fraction to the 93~GHz continuum. 

\edit1{In estimating the free-free fractions, we enforce that only two contributions of emission type, free-free and synchrotron, or, free-free and dust, compose the flux at 93~GHz. If we consider three emission types, we would obtain non-unique solutions in estimating the fractions of emission from the spectral index. Assuming two types is the best that we have assessed with the currently available data. Small dust opacities and a majority of negative slopes measured at 93~GHz independently indicate that dust does not substantially contribute in the majority of sources. Therefore assuming two dominant components to the emission at 93~GHz appears to be valid in the majority of sources. In order to decompose the emission at 93~GHz through SED fitting, additional observations at intermediate frequencies (within 2.3 GHz and 350 GHz) of comparable resolution are needed.}

\edit1{ To estimate the free-free fraction, we also assume fixed energy distribution slopes for emission from free-free, synchrotron and dust. However, variations from source to source may be expected. If we re-derive the free-free fractions, letting $\alpha_{\mathrm{d}} = 2$ without changing the free-free and synchrotron indices, we find the median dust fraction of sources remains at $f_{\mathrm{d}} = 0$. This is another indication that the spectral index assumed for dust does not have a considerable impact on the majority of sources. The range of reasonable values to consider for the frequency dependence of non-thermal emission is less constrained, where single-injection indices of $\alpha_{\mathrm{syn}} = -0.5$ to $-0.8$ can be expected all the way up to the dramatic exponential cutoff\footnote{where $S \propto \mathrm{e}^{ - \nu / \nu_b}$ and $\nu_b$ is the break frequency} in a single-injection scenario where the highest energy electrons (at the high frequency end) completely depopulate after a characteristic energy-loss time \citep[e.g.,][]{Klein2018a}. As we note in Section~\ref{sec:freefreefrac}, our assumed index of $\alpha_{\mathrm{syn}} -1.5$ agrees well with the measured value of $-1.4$ at lower resolution \citep{Bendo2016}. If we re-derive the free-free fractions assuming the canonical value of $\alpha_{\mathrm{syn}} = -0.8$ determined at 10~GHz \citep{Niklas1997}, the median free-free fraction of sources would be $f_{\mathrm{ff}} = 0.28$ and the synchrotron fraction would be $f_{\mathrm{syn}} = 0.72$. }

\edit1{On average the fractional contributions to the 93~GHz emission that we obtain assuming indices of $\alpha_{\mathrm{ff}} = -0.12$, $\alpha_{\mathrm{syn}} = -1.5$, and $\alpha_{\mathrm{d}} = 4$ appear to be reasonable and consistent. Our median free-free fraction of $f_{\mathrm{ff}} = 0.62$ with median absolute deviation of 0.29 agrees within error to the free-free fraction derived by \citep{Bendo2016} of $f_{\mathrm{ff}} = 0.84 \pm 0.10$ at 86~GHz. Our median free-free fraction is also consistent with that derived for NGC~253, $f_{\mathrm{ff}} = 0.70 \pm 0.10$, at our same frequency but averaged over 30\arcsec\ \citep{Bendo2015}. Similarly, pilot survey results of the MUSTANG Galactic Plane Survey at 90~GHz and with parsec-scale resolution indicate that $>80$\% of the emission in (candidate) clusters is composed of synchotron or free-free emission \citep{Ginsburg2020}. }

\edit1{We can also perform another self-consistent check on the free-free emission at 93~GHz, by using the recombination line emission and assuming the candidate star clusters have uniform temperatures. A constant temperature implies a constant integrated recombination line to continuum ratio (see Equation~\ref{eq:temp}). Letting $R_{\mathrm{LC}} \approx 35$~\kms\ (for a temperature of $T_e = 6000$~K), we plug in our measured line and continuum fluxes and solve for $f_{\mathrm{ff}}$. A median free-free fraction of $f_{\mathrm{ff}} = 0.85$ with median absolute deviation of 0.33 is found when considering all sources with detected recombination line emission. In comparison, the relation assumed for the in-band spectral index, also at 0.2\arcsec\ resolution, results in a median value of $f_{\mathrm{ff}} = 0.66$ and a median absolute deviation of 0.24. These values are in reasonable agreement given the uncertainties in the two methods.}

Overall, the assumptions \edit1{made in decomposing the emission at 93~GHz}, especially regarding synchrotron emissions, likely affect individual clusters at the $\sim 30\%$ level. This is not enough to bias our overall results, but follow up observations at other frequencies would be extremely helpful.

The second \edit1{potentially major source of} uncertainty is the assumed age of the burst. We discuss in Section~\ref{sec:age} how we arrive at an adopted age of the clusters of $\sim$5~Myr. The assumed age has a large impact on the stellar mass inferred from the ionizing photon rate. The ionizing photon output changes substantially (by a factor of 40 from a zero age main sequence to an age of 5~Myr) as the most massive stars explode as supernova. An uncertainty of $\sim$1~Myr about an age of 5~Myr of a star cluster results in an uncertainty in the inferred stellar mass by a factor of four. This is roughly included in the 0.4~dex uncertainty, though it would represent a systematic offset.

%%%%%%%%%%%%%%%%%%%%%%%%%%%%%%%%%%%%%%%%

\subsection{Super Star Clusters}    
\label{sec:ssc}

The estimated properties (mass, size, age; Table~\ref{tab:props}) that we derive for these candidate star clusters meet the criteria for young, massive clusters \citep[e.g.,][]{PortegiesZwart2010}, and these sources can be considered super star clusters (SSCs). Stars clusters forming in high gas surface density environments may be able to acquire significant amounts of mass before feedback effects (likely radiation pressure) can disrupt and/or disperse the cluster \citep[e.g.,][]{Adamo2016}. Super star clusters with stellar masses of $M_{\star} \gtrsim 10^5$~\Msun\ typically have high star formation efficiencies $(\varepsilon > 0.5)$ and therefore remain bound. 

With the mass surface densities that we estimate and the age of the burst equaling many multiples of the free-fall times, these star clusters will likely remain bound, at least initially. They are being born into a violent environment and clusters often still experience significant mortality after forming. Estimates of the virial mass, escape velocity and momentum driving determined through molecular line observations will help quantity their mass and initial dynamical state. 

%%%%%%%%%%%%%%%%%%%%%%%%%%%%%%%%%%%%%%%%

\subsection{Cluster Mass Function} 
\label{sec:cmf}

In Figure~\ref{fig:cmf}, we plot the cluster stellar-mass function of candidate SSCs in \trg\ and compare it with the cluster mass distributions in additional galaxies with SSC populations. 

A power-law fit to the cluster mass distribution, of 22 sources down to $M_{\star} = 2.4 \times 10^5$~\Msun,  in \trg\ results in a slope of $\beta = -1.8 \pm 0.4$. Note that the fit to the data results in $\beta = -1.76 \pm 0.07$, but given the uncertainty in our mass estimates (0.4~dex), we adopt the \edit1{more} conservative figure. We include the SSCs of NGC~253, with stellar mass properties determined on similar spatial scales ($\sim$2~pc resolution) and through H40$\alpha$ recombination lines for a zero-age main sequence population \citep{Leroy2018,Mills2020}. A power-law fit to the cluster mass distribution of NGC~253, including 12 star clusters down to $M_{\star} = 7.9\times 10^4$~\Msun, results in a slope of $\beta = -1.6 \pm 0.3$. 
We estimated the completeness limits by eye. The turn-off from power-law distributions likely includes non-physical effects, resulting from the depth/sensitivity of the observations as well as source confusion due to the high inclination in which we view the starburst regions (and which appears to be higher in NGC~253).

We also include recent results from a homogeneous analysis by \cite{Mok2020} of the cluster mass distributions of young ($\tau < 10$~Myr), massive clusters in six galaxies. As representative examples, we include the best fit power-laws from three systems: the Large Magellanic Clouds (LMC), M~51, and the Antennae System. A main result from \cite{Mok2020} is that the cluster mass functions across the six galaxies are consistent with slopes of $\beta = -2.0 \pm 0.3$. Our measurements for \trg\ and NGC~253 are consistent with these findings.

As we discuss in Section~\ref{sec:sfr}, we estimate the total stellar mass in the burst to be $8.5 \times 10^{7}$~\Msun. Extending the best fit of our cluster mass function down to a cluster mass of $\sim 2 \times 10^4$~\Msun\ would account for the additional mass and correspondingly the additional ionizing photon luminosity. As we discuss in Section~\ref{sec:dig}, if these lower mass clusters are present they would need to be located in a more-extended region than the super star clusters. Therefore the cluster mass distribution may not reach down to $2 \times 10^4$~\Msun\ in the region where the SSCs are located, but perhaps $\sim 4 \times 10^4$~\Msun\ if a third of the additional recombination line emission results from diffuse ionized gas.

%%%%%%%%%%%%%%%%%%%%%%%%%%%%%%%%%%%%%%%%

\subsection{Ionizing Photons and Diffuse Ionized Gas}    
\label{sec:dig}

\textit{Total ionizing photon rate of the starburst.} 
We measure an H40$\alpha$ recombination line flux of ($6.4 \pm 1.0$)~Jy~\kms\ integrated within the T2 aperture (see Figure~\ref{fig:totalap}) in the intermediate-configuration (0.7\arcsec) data. Assuming the temperature of $T_e = (6000\pm400)$~K from Section~\ref{sec:temp}, the total ionized content, which accounts for the full star-formation rate of the starburst, yields an ionizing photon rate of $Q_{\mathrm{T2}} = (3.9 \pm 0.3) \times 10^{53}$~s$^{-1}$. 

\textit{Ionizing photon rate from candidate super star clusters.}
The sum of the ionizing photon rate over all massive star clusters is $\approx 5.3 \times 10^{52}$~s$^{-1}$. The total integrated line flux of H40$\alpha$ in the extended-configuration (0.2\arcsec) data is $\int S\, \mathrm{dv} = (2.1 \pm 0.6)$~Jy~\kms\ within an aperture (T1) which covers the clusters in the starburst region. With the temperature of $T_e = (6000 \pm 400)$~K, the integrated line flux corresponds to an ionizing photon rate of $Q_{\mathrm{T1}} = (1.2 \pm 0.4) \times 10^{53}$~s$^{-1}$. Given the large uncertainties derived for individual clusters, we consider this value consistent with the sum over individual clusters. Therefore, we conclude that 20\% to 44\% of the total ionizing photons in the starburst can be attributed to the candidate super star clusters identified in \trg. 

\textit{Low mass clusters.}
The intermediate-configuration (0.7\arcsec) data are sensitive to emission on larger physical scales, but they also reach deeper sensitivities per unit area. Consequently, emission from a distribution of many compact, low mass clusters could be traced in these data, but not in the extended-configuration data. 

We do not find it likely that the deficit of line emission in the extended-configuration data can be attributed primarily to low mass clusters. Focusing only on the low resolution (0.7\arcsec) observations, the integrated line flux is greater in the larger aperture (T2) than in the aperture covering only the star clusters (T1), indicating that additional line emission originates outside of the area where massive clusters are forming. In order for low mass clusters to account for the difference, these would need to form in a more extended region than the bright SSCs that we see.

\textit{AGN and the circumnuclear disk.}
The circumnuclear disk, including the AGN, could be an origin for ionizing photons observed only on larger scales. The strongest recombination line emission is seen at low-resolution in this region yet no significant line detections are obtained towards Sources 17, 18, and 20. We test this scenario by extracting spectra from the 0.7\arcsec\ intermediate-configuration data and the 0.2\arcsec\ extended-configuration data in 7 non-overlapping apertures of 3\arcsec\ diameter consecutively distributed along the major axis of the starburst region. The ratios of the line flux in the intermediate- to extended-configuration data are all consistent within error; the ratios in the seven regions have a mean and standard deviation of $2.3 \pm 0.5$. This indicates that a deficit of line emission on the physical scales probed in the high resolution data is ubiquitous (and not unique to the region surrounding the AGN), and at most $\sim20$\% may originate from the circumnuclear disk  and AGN.

\textit{Diffuse ionized gas on large-scales.} 
Summarizing the information above, we find that up to $\sim$70\% of the total radio recombination line flux may originate on scales larger than $\sim$100~pc; this emission cannot be directly attributed to the AGN. Ionizing radiation may be escaping the immediate surroundings of massive stars and reaching larger scales. The approximate age of the burst indicates clusters have had time to shed their natal material, dust is not a significant contribution of their emission, low gas mass fractions have been determined, and the current free-fall times of the clusters indicate abundant time to expel gas. Despite these indicators, extinctions may still be considerable and patchy regions may help to leak ionizing photons. Note that if radiation is escaping from star clusters, this would also imply that the stellar masses (derived from the ionizing photon rates) are underestimated.

However, a missing 70\% of line flux in the 0.2\arcsec\ resolution data should be considered as an upper limit. Given the low signal-to-noise ratio of these lines, artifacts in both the image and spectral domains can impact the properties of the line profiles. For example, incomplete $uv$ coverage on the scales of the emission can result in incorrect deconvolution and thus systematic underestimates. Deeper observations at high resolution ($\sim$0.1\arcsec) would allow radio recombination lines to be mapped out with adequate signal-to-noise ratio and obtain sensitivities approaching that of the low-resolution (0.7\arcsec) observations.

%%%%%%%%%%%%%%%%%%%%%%%%%%%%%%%%%%%%%%%%

\subsection{Role of the AGN}    
\label{sec:agn}

Although the Seyfert 2 AGN in \trg\ is one of the brightest in our sky at X-ray energies, obscuration by e.g., $A_{\mathrm{V}} \geq 160$ \citep{Spoon2000} has prevented its observation at virtually all other wavelengths (also given the spatial resolutions observed). Our observations at 93~GHz provide another piece of evidence for its existence. 

We detect a point source in the 93~GHz image with a deconvolved size determined by \texttt{PyBDSF} of $\mathrm{FWHM} = 0.84 \pm 0.01$~pc. The flux of $9.7 \pm 1.0$~mJy that we extract from the source is $\sim10$\% of the total continuum ($\sim$120~mJy) flux in the 0.12\arcsec\ resolution data. The Band 3 spectral index we derive is $\alpha_{93} = -0.85 \pm 0.05$, which is consistent with freshly accelerated electrons emitting synchrotron radiation. However, we follow the procedure outlined in Section~\ref{sec:freefreefrac}, assuming free-free emission may also contribute, and place a limit on the ionizing photon rate. The spectral index at 93~GHz corresponds to $f_{\mathrm{ff}} \geq 0.47 \pm 0.03$. If we assume a temperature of $T_e = 6000$~(30\,000)~K this would correspond to $Q_0 = 1.1 \,(0.5) \times 10^{52}$~s$^{-1}$. The low level of escaping ionizing radiation is consistent with previous work from MIR line ratios \citep{Spoon2000}. 

We place a limit on the ionizing photons that could be leaking into medium. The bolometric luminosity estimated from the X-ray luminosity is  $L_{\mathrm{bol}} \approx 6 \times 10^{43}$~erg~s$^{-1}$, which we determined using the relation for Seyfert AGN of $L_{\mathrm{bol}} \sim 20 L_{\mathrm{X}}$ \citep{Hopkins2007}. Given the bolometric luminosity, we can now estimate the total expected ionizing photon luminosity. We use the standard relation of \cite{Elvis1994}, which posits $L_{\mathrm{bol}} / L_{\mathrm{ion}} \sim 3$ with an average ionizing photon energy of 113~eV. Therefore the expected ionizing photon luminosity of the AGN is $L_{\mathrm{ion}} \approx 2 \times 10^{43}$~erg~s$^{-1} \approx 5 \times 10^{9}$~\Lsun, while our limit corresponds to $\approx 2 \times 10^{42}$~erg~s$^{-1}$. This would imply that $<10$\% of the ionizing photon luminosity of the AGN is escaping into the surrounding nuclear starburst and creating ionized gas.

%%%%%%%%%%%%%%%%%%%%%%%%%%%%%%%%%%%%%%%%

\subsection{Total Burst of Star-formation}    
\label{sec:sfr}

We can convert the total ionizing photon rate of the burst $Q_{\mathrm{T2}}$ (see Section~\ref{sec:dig}), as measured in the T2 region, to a luminosity. Assuming an average ionizing photon energy of $\left<h\nu\right> \approx 17$~eV, the ionizing luminosity is $L_{\mathrm{T2}} \approx 1.1 \times 10^{43}$~erg~s$^{-1} = 2.8 \times 10^{9}$~\Lsun. A ratio of bolometric luminosity to ionizing photon luminosity $L_{\mathrm{bol}}$/$L_0 \sim 14$ is predicted for a 5~Myr old population, using Starburst99 \citep{Leitherer1999}. Thus, we can expect a bolometric luminosity of $4 \times 10^{10}$~\Lsun\ which agrees within error to the bolometric luminosity derived from FIR observations, $L_{\mathrm{bol}} = (2.0 \pm 0.2) \times 10^{10}$~\Lsun\ \citep{Bendo2016}.

We note that, if we had adopted a lifetime for the burst of 4~Myr rather than 5~Myr, Starburst99 calculations expect a bolometric luminosity of $L_{\mathrm{bol}} \approx 2 \times 10^{10}$~\Lsun. 
Also, this calculation assumes the source of the additional recombination line emission in the intermediate-configuration data is also characterized by a 5~Myr old stellar population.

The ratio of bolometric luminosity to mass of a 5~Myr old burst is estimated at $\Psi \sim 470$~\Lsun/\Msun. With an $L_{\mathrm{bol}} = 4 \times 10^{10}$~\Lsun, the inferred total stellar mass of the population is $\sim 8.5 \times 10^7$~\Msun. The total stellar mass in our candidate clusters is $\approx 1.1 \times 10^7$~\Msun, or $\sim$13\% of the expected total stellar mass.   

%%%%%%%%%%%%%%%%%%%%%%%%%%%%%%%%%%%%%%%%

\subsection{Star Clusters and the Central Wind}    
\label{sec:outflow}

An outflow of warm ionized gas in \trg\ has been modeled as a biconical outflow with a deprojected velocity of $\mathrm{v} \approx 525$~\kms\ with emission out to 1.8~kpc \citep{Heckman1990}. The outflow has been traced in [NII], [SII] and H$\alpha$ in multiple analyses \citep{Heckman1990, Moorwood1996, Mingozzi2019}. \cite{Heckman1990} estimated a total energy of $2 \times 10^{55}$~erg and momentum flux of $9 \times 10^{33}$~dyne~$= 1400$~\Msun~\kms~yr$^{-1}$ (after correcting for the updated distance). These values do not include possible contributions by colder phases, which are likely associated with the \trg\ outflow (Bolatto et al., in prep.). While these values are uncertain, they are useful for an order of magnitude comparison with expected properties of the star clusters. 

Gas reaching 1.8~kpc and moving at 525 \kms\ would have been launched $\sim$3~Myr ago assuming a constant velocity, which is within the time-frame of the burst. A $\sim$3~Myr time-scale and a velocity of 525 \kms\ would put $8 \times 10^6$~\Msun\ of warm ionized gas mass into the outflow.
Using Starburst99 \citep{Leitherer1999}, an estimated mechanical luminosity of $\sim 3 \times 10^{41}$~erg~s$^{-1}$ from the clusters, fairly constant over the 5~Myr age, equates to an injected energy of $\approx6 \times 10^{55}$~erg. It would require 30\% of the expected mechanical energy output of the clusters to drive the ionized outflow. From simulations, the total momentum supplied to the ISM per supernova is expected to be $2.8 \times 10^5$~\Msun~\kms\ \citep[][and references therein]{Kim2015}, with stellar winds contributing an additional 50\%  to the momentum. Starburst99 predicts the supernova rate to be fairly constant at $\sim$0.009~yr$^{-1}$ for these clusters. From these values, we estimate that the momentum supplied to the ISM by the candidate star clusters would be $\sim3800$~\Msun~\kms~yr$^{-1}$. It would require about 40\% of the expected momentum output of the clusters to drive the ionized outflow. Therefore, neither the energetics nor the inferred momentum of the outflow prevent it from being driven solely by the star clusters in \trg, although there are very considerable uncertainties associated with this calculation.

%%%%%%%%%%%%%%%%%%%%%%%%%%%%%%%%%%%%%%%%

\subsection{Comparison with NGC~253}    
\label{sec:ngc253}

NGC~253 is a nearly edge-on galaxy located nearby \citep[$3.5\pm 0.2$~Mpc ;][]{Rekola2005} with similar properties as \trg. It hosts a central starburst spanning $\sim$200~pc with young, massive clusters \citep{Leroy2018,Mills2020}. A major difference between these two galaxies is that \trg\ shows \edit1{unambiguous} signatures of harboring an active super-massive black hole. Since \trg\ is just the second analysis we have undertaken with millimeter wavelength ALMA observations on scales which resolve the clusters, it is relevant to compare the properties of their super stars clusters.

Analyses of the cluster population of NGC~253 at $\sim$2~pc resolution with ALMA \citep{Leroy2018,Mills2020} revealed 18 sources as (proto-)super star clusters which are in the process of forming or close to a zero age main sequence ($\sim$1~Myr) \citep[see also ][]{Rico-Villas2020}. Overall the properties of the clusters are strikingly similar as those in \trg; they have a median stellar mass of $1.4\times10^{5}$~\Msun\ and FWHM sizes of 2.5--4~pc. The star clusters of NGC~253 boast slightly higher ionizing photon luminosities and larger gas fractions $M_{\mathrm{gas}} / M_{\mathrm{tot}} \sim 0.5$, reflecting a slightly younger age. The slopes of the cluster mass function we derive are also consistent (see Section~\ref{sec:cmf}), just offset by a factor of $\sim$3.5 in mass. As in \trg, at least 30\% of the of the nuclear starburst of NGC~253 originates in a clustered mode of star-formation.

While we have not estimated some of the dynamical properties of \trg, the presence of broad recombination line emission in four sources in NGC~253 indicates the star clusters are operating under similar processes. In NGC~253, the sources are young enough that feedback has not managed to unbind a large fraction of the gas from the clusters. The slightly higher total mass surface densities and smaller free-fall times in \trg\ indicate its clusters might be surviving a young, disruptive stage. 

The concerted feedback of the young ($<10$~Myr), forming SSCs identified in NGC~253 \citep{Leroy2018} are likely not responsible for the starburst driven outflow, \edit1{as they are expected to impart momentum that is a factor of 10-100 lower than the outflow momentum estimated in CO} \citep{Bolatto2013,Krieger2019}. Some (global) event may be initiating the active formation of star clusters. On the other hand, the star clusters in \trg\ could potentially influence the outflow of warm ionized gas; some event appears to be inhibiting the formation of new star clusters.

%%%%%%%%%%%%%%%%%%%%%%%%%%%%%%%%%%%%%%%%

\section{Summary}   
\label{sec:conclude}

Massive, clustered star-formation is an efficient and possibly common mode of star-formation in high gas density environments. Nearby galaxies with bursts of star-formation in the central $\mathcal{O}$(100~pc) are local laboratories to study this mode, and for \trg, in the presence of a Seyfert AGN. High levels of dust extinction ($A_V \sim 40$) in the nearly edge-on ($i \sim 72^{\circ}$) central region of \trg\ have previously prevented the direct observation and characterization of its massive star clusters.

We identify 27 super star cluster candidates in the central starburst of \trg. We derive properties of the candidate clusters through ALMA observations at 2.2~pc resolution of the 93~GHz (3~mm) free-free emission and hydrogen recombination line emission (H40$\alpha$ and H42$\alpha$) arising in photo-ionized gas. 
We also use and present high-resolution 350~GHz imaging of the dust continuum observed with ALMA, and supplement our analysis with 2.3~GHz continuum imaging which primarily traces synchrotron emission \citep{Lenc2009}. 

Our results are as follows:
\begin{itemize}
    \item The 27 point sources identified  ($ \geq10\sigma$) in 93~GHz continuum emission as candidate super star clusters have FWHM sizes of 1.4--4.0~pc.
    The 93~GHz emission in these bright, compact regions is dominated by free-free emission, with a median free-free fraction of $f_{\mathrm{ff}} = 0.62$. Synchrotron emission from recent supernova remnants contributes to the 93~GHz emission with a median fraction of $f_{\mathrm{syn}} = 0.36$. Substantial dust emission is found in three sources. 
    
    \item We average the spectra of the H40$\alpha$ and H42$\alpha$ recombination lines to synthesize an effective H41$\alpha$ profile. 
    Recombination line emission is detected in 15 candidate clusters, generally with narrow (FWHM~$\sim36$~\kms) line widths. Six of the detected sources have significant ($f_{\mathrm{syn}} \gtrsim 0.5$) synchrotron emission; three of those have broad line widths with FWHM~$> 105$~\kms.

    \item We estimate an electron temperature of $T_e = (6000 \pm 400)$~K of the ionized gas using the flux ratio of the integrated line to free-free continuum. This electron temperature implies an average metallicity of $12 + \log_{10}(\mathrm{O/H}) = 8.9 \pm 0.1$ surrounding these young massive stars. The ionized gas densities of $\log_{10} (n_e/\mathrm{cm}^{-3})= 3.1$--3.9, that we derive are typical of classic H~II regions. The ionized gas masses of the clusters are typically $<1$\% of the estimated stellar mass. 

    \item We determine ionizing photon rates of the candidate SSCs in the range $\log_{10} (Q_0 / \mathrm{s}^{-1}) \sim 50.4$ -- 51.8. Adopting an age of $\sim$5~Myr, the stellar masses implied by the ionizing photon rates are $\log_{10} ( M_{\star} / \mathrm{M_{\odot}} ) \sim$~4.7--6.1. The sum of the stellar masses of the candidate SSCs is $\approx 1 \times 10^7$~\Msun. The uncertainties on these measurements are 0.4~dex. We discuss the age estimate in Section~\ref{sec:age}.

    \item We fit the cluster stellar-mass distribution and find a slope of $\beta = -1.8 \pm 0.4$. The slope of the fit is consistent with our fit to the candidate SSCs in the central starburst of NGC~253 \citep{Mills2020} and to recent findings in the LMC, M~51, and the Antennae system \citep{Mok2020}. 

    \item We estimate the gas mass of the candidate clusters from dust emission. The total mass, $M_{\mathrm{Tot}}$, is evaluated by the combined stellar and gas masses. Gas mass fractions range from $M_{\mathrm{gas}}/M_{\mathrm{Tot}}=0.04$--0.22, with a median value of 0.13. We calculate the total-mass surface density of the clusters and find a median value of $\Sigma_{\mathrm{Tot}} = 3 \times 10^{4}$~\Msun~pc$^{-2}$. The median free-fall timescale is $0.04$~Myr. 

    \item With low-resolution (0.7\arcsec) observations of the H40$\alpha$ recombination line, we measure a total ionizing photon rate of the burst of $Q_0 = (3.9 \pm 0.3) \times 10^{53}$~s$^{-1}$. The candidate star clusters that we analyze contribute 20--44\% of the total ionizing photon rate. Additional recombination line emission present in the low-resolution data appears to be ubiquitous throughout the starburst region, cannot be directly attributed to the AGN, and is also found outside of the area where the SSCs are located. Diffuse ionized gas may be responsible for some of the additional emission, although low mass clusters and distributed star formation are expected to also contribute. This indicates that ionizing radiation may be escaping the immediate surroundings of massive stars and reaching the larger ($>100$~pc) scales traced at lower resolution. 

    \item We compare the candidate SSCs in \trg\ with the (proto-)SSCs recently identified in the central starburst of NGC~253 \citep{Leroy2018,Mills2020}. The age of the burst (1~Myr) is a bit younger in NGC~253, the gas mass fractions ($\sim$0.5) are a bit higher, and the stellar mass of the clusters are slightly smaller (factor of 3.5). While the actively forming clusters are not a major contributor to the starburst driven outflow in NGC~253, the slightly older population of the star clusters in NGC~4945 may contribute to driving a nuclear outflow of warm ionized gas. 
    \item Strong, variable X-ray emission, which is Compton thick, provides evidence for a Seyfert AGN in \trg. The bright, 93~GHz source which we presume to be the AGN is point-like in our 2.2~pc beam. We measure its in-band spectral index at 93~GHz to be $\alpha_{93} = -0.85 \pm 0.05$, likely dominated by synchrotron emission.  We do not detect recombination line emission from this point source. Our observations support previous findings in which UV ionizing radiation from the AGN is heavily obscured in all directions. We estimate an upper limit for its escaping ionizing photon rate of $Q_0 < 1 \times 10^{52}$~s$^{-1}$, which is $\lesssim 10$\% of the expected luminosity of ionizing photons for a typical Seyfert AGN.
    
    \item Lastly, we report on a shortcoming of H42$\alpha$ as a ``low-resolution'' tracer of ionizing radiation. When observed with low spatial resolution and/or from broad line components, we find this spectral line is likely contaminated by line emission from other species, resulting in the recombination line flux (and star-formation rates derived from it) being overestimated by a factor of $\sim$2, when compared to a similar analysis of H40$\alpha$. 

\end{itemize}

%%%%%%%%%%%%%%%%%%%%%%%%%%%%%%%%%%%%%%%%

\acknowledgments

The authors thank the anonymous referee for the careful review of the article and helpful input. We also thank Emil Lenc for providing the Australian LBA 2.3~GHz data and Paul van der Werf for the HST Pa-$\alpha$ data.

KLE acknowledges financial support from the Netherlands Organization for Scientific Research (NWO) through TOP grant 614.001.351. KLE thanks the Laboratory for Millimeter Wave Astronomy and the Department of Astronomy at the University of Maryland and the Green Bank Observatory for hosting her during completion of this work. ADB and RCL acknowledge support from NSF through grants AST-1412419 and AST-1615960.  AKL acknowledges support by the National Science Foundation (NSF) under Grants No.1615105, 1615109, and 1653300, as well as by the National Aeronautics and Space Administration (NASA) under ADAP grants NNX16AF48G and NNX17AF39G. EACM gratefully acknowledges support by the National Science Foundation under grant No. AST-1813765. ER acknowledges the support of the Natural Sciences and Engineering Research Council of Canada (NSERC), funding reference number RGPIN-2017-03987.

This article makes use of the following ALMA data: ADS/JAO.ALMA\#2018.1.01236.S, and ADS/JAO.ALMA\#2016.1.01135.S. ALMA is a partnership of ESO (representing its member states), NSF (USA) and NINS (Japan), together with NRC (Canada), NSC and ASIAA (Taiwan), and KASI (Republic of Korea), in cooperation with the Republic of Chile. The Joint ALMA Observatory is operated by ESO, AUI/NRAO and NAOJ. The National Radio Astronomy Observatory is a facility of the National Science Foundation operated under cooperative agreement by Associated Universities, Inc.


\facilities{ALMA, Australian LBA, HST}


\software{  APLpy \citep{Robitaille2012}, 
            Astropy \citep{Astropy2018},
            CASA \citep{McMullin2007},
            CRRLpy \citep{Salas2016},
            Matplotlib \citep{Hunter2007},
            PyBDSF \citep{Mohan2015}.
          }

\bibliography{papers-ngc4945}{}
\bibliographystyle{aasjournal}

%%%%%%%%%%%%%%%%%%%%%%%%%%%%%%%%%%%%%%%%


\appendix

\section{Millimeter wavelength emission from free-free continuum and recombination lines}
\label{ap:derivation}

In this section we present the relations for the recombination line intensity and free-free continuum applicable for millimeter wavelength emission, largely by bringing together information derived in \cite{Gordon2009} and \cite{Draine2011}. We show how these relations can be used to estimate the temperature and ionizing photon rate of the emitting plasma.

At frequencies of $\mathcal{O}$(10 GHz), the electron temperature $T_e$ of the thermal emission from ionized gas accurately characterizes the relative populations of electrons in bound atomic levels, and the system is well approximated by local thermodynamic equilibrium (LTE). Collisions dominate and line emission is described by the Boltzmann distribution.

As the collisional cross section of a Bohr atom rapidly decreases ($\propto \mathsf{n}^{4}$) towards smaller principal quantum numbers, or higher frequencies of $\mathcal{O}$(100 GHz), collisions become less important, and the observed line intensity is not exactly set by the kinetic motion of the electrons. Radiative processes, which dominate the smallest principal quantum numbers, influence the level populations; smaller $\mathsf{n}$ are underpopulated as compared with a Boltzmann distribution. The population rates into an energy level are not exactly balanced by rates out of the level, and a correction coefficient ($b_{\mathsf{n}}$) for a departure from LTE is necessary.

%%%%%%%%%%%%%%%%%%%%%%%%%%%%%%%%%%%%%%%%

\subsection{Recombination Line Intensity}   
\label{ap:line}

Considering Kirchoff's law of thermodynamics with no net change in intensity through the medium, the emission of a radio recombination line in LTE, $S_{\mathsf{n}} (\mathrm{LTE})$, is related to $\kappa_{\mathsf{n}}$, the fractional absorption per unit pathlength $\ell$, for small line optical depths by
\begin{equation}
    S_{\mathsf{n}} (\mathrm{LTE}) \approx \kappa_{\mathsf{n}} \, \ell \, B_{\nu}(T_e) \, \Omega
\end{equation}
for a transition to final principal quantum number $\mathsf{n}$, where $B_{\nu}$ is the Planck function and $\Omega$ is the solid angle on the sky.
For small optical depths of the line and the free-free continuum $\tau_{\mathrm{c}}$, the non-LTE line flux density is, by definition, 
\begin{equation}
    S_{\mathsf{n}} \approx S_{\mathsf{n}}(LTE) b_{\mathsf{n+1}}  \left( 1 + \frac{\tau_C}{2} \beta \right).
\end{equation}
Here, $b_{\mathsf{n+1}}$ is the departure coefficient, which is defined as $b_{\mathsf{n}} = n / n_{\mathrm{LTE}}$, the ratio of the actual number density of atoms with an electron in level $\mathsf{n}$ to the number which would be there if the population were in LTE at the temperature of the ionized gas, such that for LTE, $b_{\mathsf{n}} = 1$. $\beta$ is the departure coefficient which accounts for stimulated emission. The second term in parenthesis is usually negligible at millimeter wavelengths, such that $(1 + 0.5\tau_c \beta) \approx 1$. In Eq.~\ref{aeq:kappac} of Section~\ref{ap:cont}, the absorption coefficient of the free-free continuum is given and the optical depth can be derived to show that typical parameters (e.g. $EM_l = 10^7$~cm$^{-6}$~pc and $T_e = 10^4$ K) of ionized regions around massive stars result in $\tau_c \lesssim 3 \times 10^{-6}$ at 100~GHz. Typical values of $\beta$ for $\mathsf{n} \gtrsim 40$ are $ \left| \beta \right| \lesssim 40$ \citep[e.g.,][]{Storey1995}.

In its expanded form, the LTE absorption coefficient of a radio recombination line can be expressed in terms of the Saha-Boltzmann distribution, oscillator strength, and line frequency as 
\begin{equation}
    \kappa_{\mathsf{n}} \approx 
    n_e n_+ \frac{2}{\sqrt{\pi}} \frac{e^2}{m_e} 
    \left( \frac{ h^2}{2 m_e k_B }\right)^{3/2} 
    \frac{h}{k_B }~T_e^{-2.5} R_H Z^2 \Delta \mathsf{n} 
    \left( 1 - \frac{3}{2} \frac{\Delta \mathsf{n}}{\mathsf{n}} \right) 
    \exp\left( \frac{ \chi_{\mathsf{n}} }{ k_B T_e} \right)
    M(\Delta \mathsf{n}) 
    \left( 1 + \frac{3}{2} \frac{\Delta \mathsf{n}}{\mathsf{n}} \right)
    \phi_{\nu}
\end{equation}
where $n_e$ and $n_+$ are the number densities of electrons and ions respectively; $R_H$ is the Rydberg constant for hydrogen; $Z$ is the effective nuclear charge; $\Delta \mathsf{n}$ is the change in energy levels of the given transition; $\chi_{\mathsf{n}}$ is the energy required to ionized the atom from state $\mathsf{n}$, but $\exp\left(\chi_{\mathsf{n}}/ k_{B}T_e \right)$ is small $(<1.02)$ for $\mathsf{n} \geq $ 40  and typical ionized gas temperatures; $M(\Delta \mathsf{n} = 1,2) = 0.190775, 0.026332$ is an approximation factor for the oscillator strength; and, $\phi_{\nu}$ is the line profile (normalization; in SI units, Hz$^{-1}$) such that $\int_{-\infty}^{\infty} \phi_{\nu}~\mathrm{d}\nu = 1$.

Bringing the above equations together, using the Rayleigh-Jeans approximation of the Planck function, and integrating over the line profile, we arrive at the expression for the integrated line flux density, 
\begin{equation}
    \int S_{\mathsf{n}} \, \mathrm{dv} =  
	\left( 65.13~\mathrm{mJy~km~s^{-1}}\right) 
 	b_{\mathsf{n+1}} 
	\left( \frac{n_e n_p V }{ 5\times10^8~\mathrm{cm^{-6}~pc^3} }\right) 
	\left( \frac{ D }{ 3.8~\mathrm{Mpc} }\right)^{-2}
	\left( \frac{ T_e }{ 10^4~\mathrm{K} } \right)^{-1.5} 
	\left( \frac{\nu}{100~\mathrm{GHz}} \right)
\end{equation}
where we take $n_+ = n_p$, $Z = 1$, and $\Delta \mathsf{n} = 1$. For convenience and self consistency, we  express the emission measure in terms of source volume, $V$, and the solid angle in terms of distance to the source, $D$. Let $\Omega = \frac{ \pi r^2 }{D^2}$, where $r$ is the radius of the region, and if we let $\ell = \frac{4}{3} r$, the volumetric emission measure is given as $EM_{V} = n_e n_p V$ where $V = \frac{4}{3} \pi r^3$.

%%%%%%%%%%%%%%%%%%%%%%%%%%%%%%%%%%%%%%%%

\subsection{Continuum Intensity} 
\label{ap:cont}

The absorption coefficient for free-free continuum \citep{Oster1961} in the Rayleigh-Jeans limit is, as a function of frequency,
\begin{equation}
    \kappa_{\mathrm{c}} =  \frac{ n_e n_+ }{ \nu^2 } \frac{ 8 Z^2 e^6 }{3 \sqrt{3} m_e^3 c} \left( \frac{ \pi }{ 2 } \right)^{1/2} \left( \frac{m_e}{k_B T_e} \right)^{3/2} g_{\mathrm{ff}} 
\label{aeq:kappac}
\end{equation}
where the gaunt free-free factor is \citep{Draine2011}
\begin{equation}
    g_{\mathrm{ff}} \approx 13.91 \left( Z \frac{\nu}{ \mathrm{Hz} } \right)^{-0.118} \left( \frac{T_e}{\mathrm{K}} \right)^{0.177}
\end{equation}
valid for  $ \nu_p \ll \nu \ll kT_e/ h$ where the plasma frequency is $\nu_p = 8.98(n_e /\mathrm{cm}^{-3})^{1/2}$~kHz, and to within 10\% when $1.4 \times 10^{-4} < Z \nu T_e^{-1.5} < 0.25$.

The continuum intensity for an optically thin medium is
\begin{equation}
    S_{\mathrm{c}} \approx B_{\nu} \, \kappa_{\mathrm{c}} \, \ell \, \Omega .
\end{equation}
As we did for the line intensity, we can also express the above relation in terms of distance $D$ and the volumetric emission measure of the region $EM_V = n_e n_p V$ where $V = \frac{4}{3} \pi r^3$:
\begin{equation}
    S_{c} = (2.080~\mathrm{mJy} ) 
    \, Z^{1.882}
	\left( \frac{ n_e n_+ V }{ 5\times10^8~\mathrm{cm^{-6}~pc^3} } \right) 
	\left( \frac{ D }{ 3.8~\mathrm{Mpc} } \right)^{-2} 
	\left(  \frac{T_e}{ 10^4~\mathrm{K}} \right)^{-0.323} 
	\left( \frac{\nu}{ 100~\mathrm{GHz} } \right)^{-0.118} 
\label{aeq:cont}
\end{equation}
evaluated in the Rayleigh-Jeans limit. 

%%%%%%%%%%%%%%%%%%%%%%%%%%%%%%%%%%%%%%%%

\subsection{Physical Properties}   
\label{ap:props}

%%%%%%%%%%%%%%%%%%%%%%%%%%%%%%%%%%%%%%%%

\subsubsection{Line to Continuum Ratio and Temperature}   
\label{ap:temp}

Using the previous expressions, the ratio between the RRL and continuum flux density is:
\begin{equation}
\frac{ \int S_{\mathsf{n}}~\mathrm{dv} }{S_{\mathrm{c}}} = 
    \left( 31.31~\mathrm{km~s^{-1}} \right) 
 	b_{\mathsf{n+1}} 
	\left( \frac{ T_e }{ 10^4~\mathrm{K} } \right)^{-1.177} 
	\left( 1 + y \right)^{-1}
	\left( \frac{\nu}{100~\mathrm{GHz}} \right)^{1.118}
\label{aeq:lc_ratio}
\end{equation}
where $ y = n_{\mathrm{He^+}} / n_p$ is the ratio of singly ionized helium to hydrogen by number and $n_{\mathrm{He^+}}$ is the singly ionized helium number density. If we assume the emission region is composed of only hydrogen and singly ionized helium, then
\begin{equation}
    \frac{ n_p }{ n_+} = \frac{ n_p }{ n_p + n_{\mathrm{He^+}} } = \left(  1 + \frac{ n_{\mathrm{He^+}} }{ n_p } \right)^{-1} = \left( 1 + y \right)^{-1}.
\end{equation}
We can rearrange the integrated RRL line to continuum ratio and solve for the electron temperature $T_e$ of the emission region,
\begin{equation}
    T_e = 10^4~\mathrm{K}
	\left[ b_{\mathsf{n+1}} 
	\left( 1 + y \right)^{-1}
	\left( \frac{ R_{\mathrm{lc}} }{ 31.31~\mathrm{km~s^{-1}} } \right)^{-1}
	\left( \frac{\nu}{100~\mathrm{GHz}} \right)^{1.118} \right]^{0.85}
\label{aeq:temp}
\end{equation}
where we denote the integrated RRL line to continuum ratio as $R_{\mathrm{lc}} = \frac{ \int S_{\mathsf{n}}~\mathrm{dv} }{ S_{\mathrm{c}} } $.

%%%%%%%%%%%%%%%%%%%%%%%%%%%%%%%%%%%%%%%%

\subsubsection{Ionizing Photon Rate}   
\label{ap:q}

The rate of ionizing photons ($E>13.6 eV$) is given by,
\begin{equation}
    Q_0 = n_e n_+ V \alpha_B
\end{equation}
where $\alpha_B$ is the case B recombination coefficient \citep{Draine2011},
\begin{equation}
    \alpha_B =  2.59 \times 10^{-13} ~\mathrm{cm^3~s^{-1}} \left( \frac{ T_e }{ 10^4~\mathrm{K} } \right)^{-0.833 - 0.034 \ln{(T_e\,/\,10^4\,\mathrm{K})} }
\end{equation}
which is valid for 3000~K~$< T_e <$~30,000~K. Thus we have,
\begin{equation}
    Q_0 = \left( 3.805 \times 10^{51}~\mathrm{s}^{-1} \right) \left( \frac{ n_e n_+ V }{ 5 \times 10^8~\mathrm{cm^{-6}~pc^3} } \right) \left( \frac{ T_e }{ 10^4~\mathrm{K} } \right)^{-0.833 - 0.034 \ln{(T_e\,/\,10^4\,\mathrm{K})} }.
\label{aeq:q}
\end{equation}

%%%%%%%%%%%%%%%%%%%%%%%%%%%%%%%%%%%%%%%%

\section{Recombination Line Spectra of all Sources}
\label{ap:spec}

In Figure~\ref{fig:ap_allspec}, we show that the central velocities of our detected recombination lines are in good agreement with the kinematic velocity expected of the disk rotation, though we include the recombination line spectra for all 29 sources (including those that are not significantly detected). The recombination line spectra from our sources are overlaid on H40$\alpha$ spectra extracted from the intermediate configuration observations (0.7\arcsec\ resolution). High-resolution recombination line emission is coincident with emission from the intermediate configuration data.

\begin{figure*}
\centering
\includegraphics[width=0.8\textwidth]{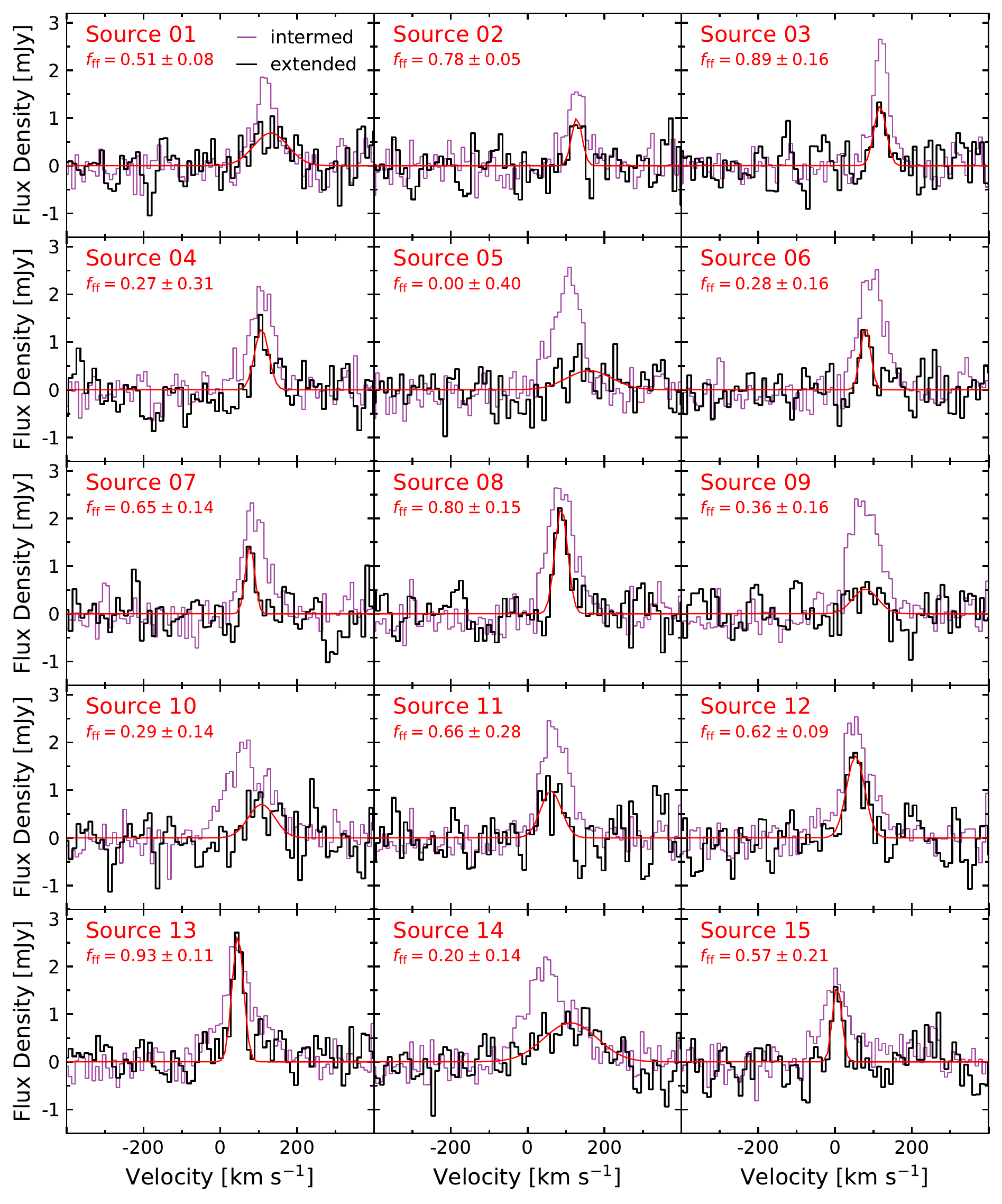}
\caption{ {\it (cont.)}}
\label{fig:ap_allspec}
\end{figure*}

\begin{figure*}
\figurenum{12}
\centering
\includegraphics[width=0.8\textwidth]{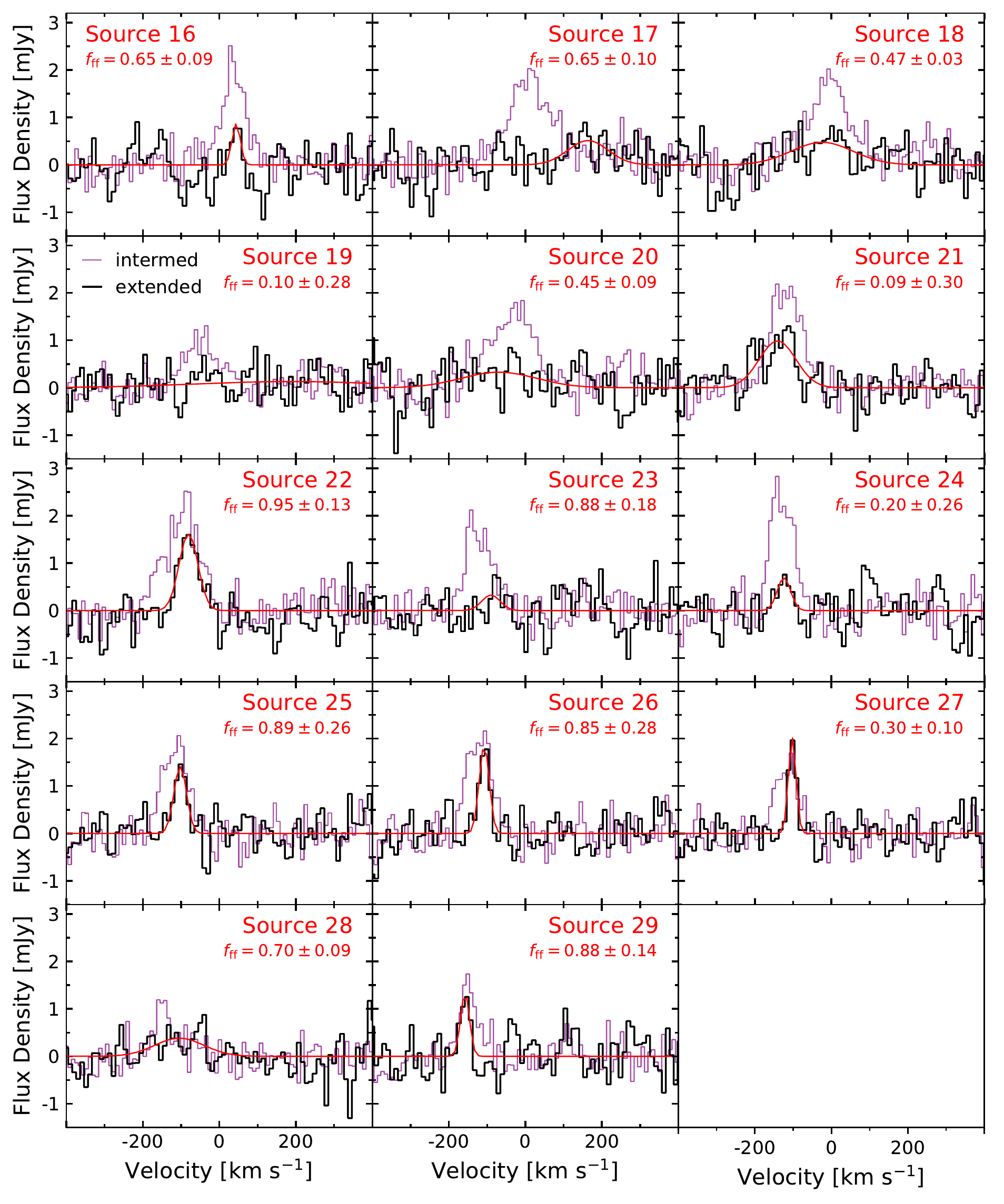}
\caption{ Effective H41$\alpha$ recombination line spectra (thick, black line) for sources with significantly detected emission and the best fit line profile (red) -- same as in Figure~\ref{fig:spectra}. H40$\alpha$ line emission tracing the kinematics is overlaid in thin purple; it represents the H40$\alpha$ spectrum centered at the same source locations and extracted from low resolution ($\sim$0.7\arcsec) observations.}
\end{figure*}

%%%%%%%%%%%%%%%%%%%%%%%%%%%%%%%%%%%%%%%%

\section{Spectral Energy distribution of all Sources}
\label{ap:sed}

\begin{figure*}
\centering
\includegraphics[width=0.8\textwidth]{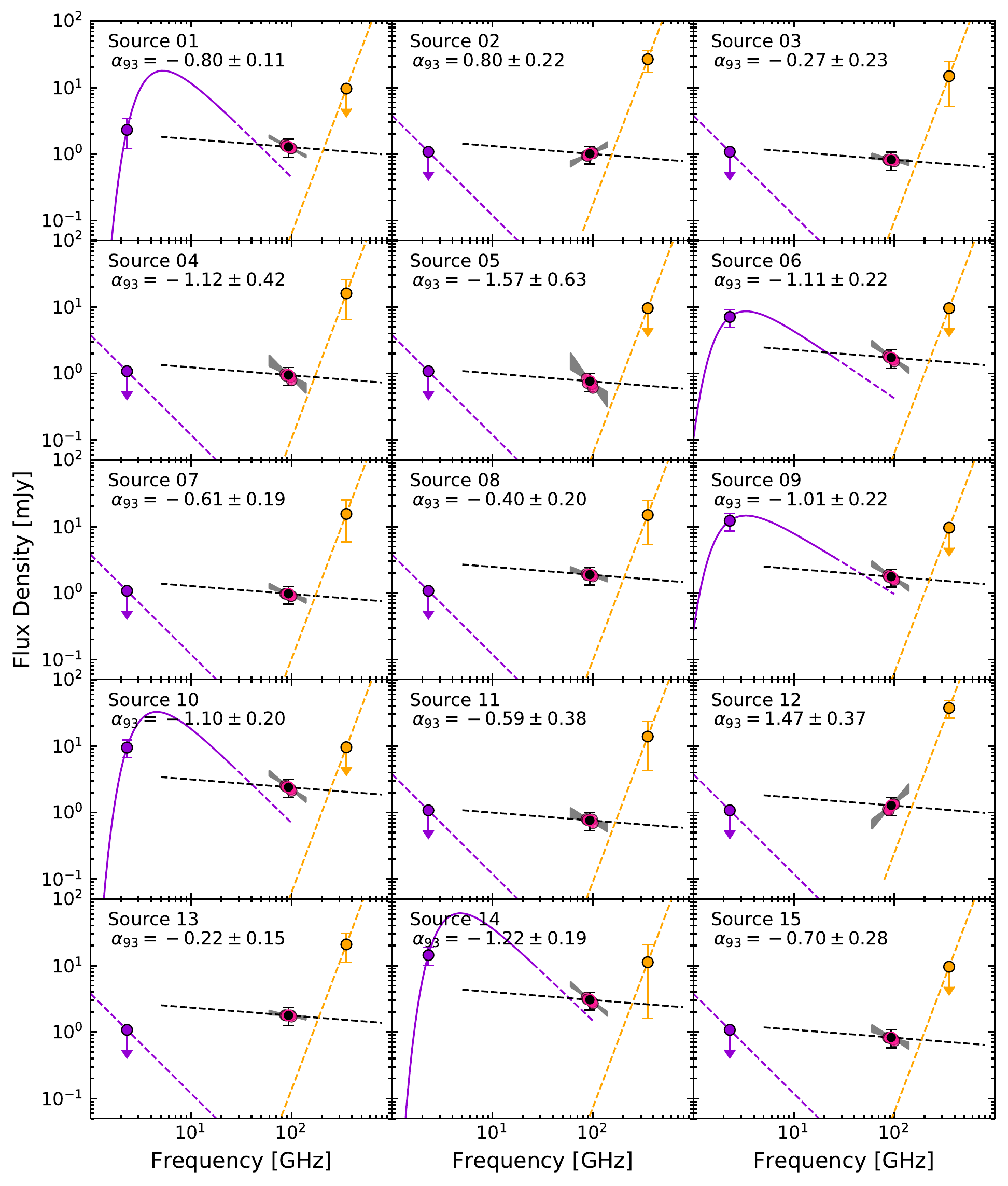}
\caption{ {\it (cont.)}}
\label{fig:ap_sed}
\end{figure*}

\begin{figure*}
\figurenum{13}
\centering
\includegraphics[width=0.8\textwidth]{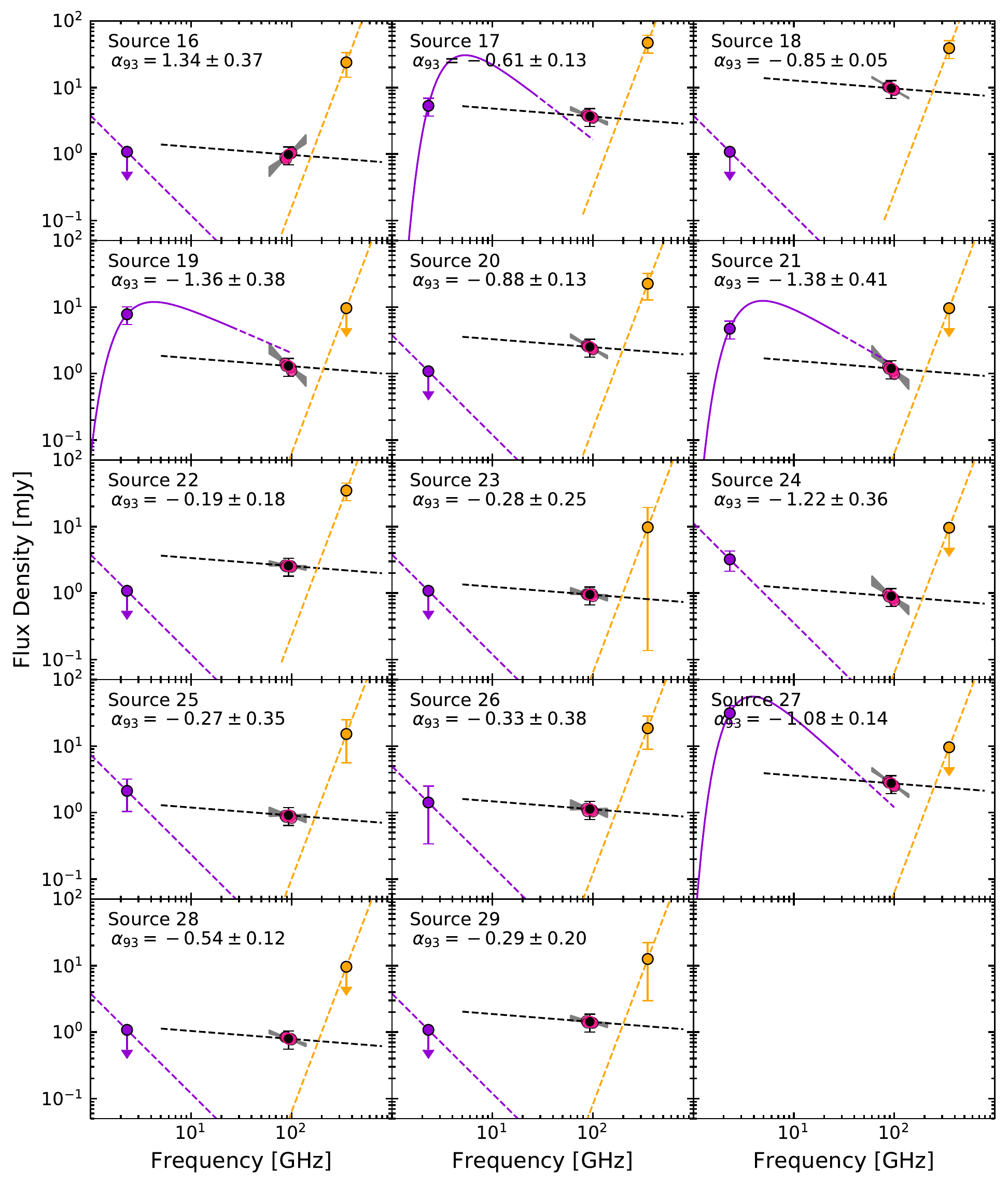}
\caption{SEDs constructed for each source, same as in Figure~\ref{fig:SED}. The dashed green line represents a dust spectral index of $\alpha = 4.0$, normalized to the flux density we extract at 350~GHz (orange data point). The dashed black line represents a free-free spectral index of $\alpha = -0.12$,  normalized to the flux density we extract at 93~GHz (black data point). The white data points show the flux densities extracted from the band 3 spectral windows. The gray shaded region is the 1$\sigma$ error range of the band 3 spectral index fit, except we have extended the fit in frequency for displaying purposes. The purple dashed line represents a synchrotron spectral index of $\alpha = -1.5$,  normalized to the flux density we extract at 2.3~GHz (purple data point); except for Source 14 where the solid purple line represents the matched 2.3--23 GHz fit.  Error bars on the flux density data points are 3$\sigma$.}
\end{figure*}

%%%%%%%%%%%%%%%%%%%%%%%%%%%%%%%%%%%%%%%%

%\bibliography{papers-ngc4945}{}
%\bibliographystyle{aasjournal}



\end{document}